\documentclass[10pt,journal,compsoc]{IEEEtran}
\ifCLASSOPTIONcompsoc
  \usepackage[nocompress]{cite}
\else
  \usepackage{cite}
\fi
\ifCLASSINFOpdf
  \usepackage[pdftex]{graphicx}
   \DeclareGraphicsExtensions{.pdf,.jpeg,.png}
\else
\fi
\usepackage{amsmath}

\usepackage{amssymb}

\usepackage{cite}

\newtheorem{theorem}{Theorem}[section]

\begin{document}
\title{CISER: An Amoebiasis inspired Model for Epidemic Message Propagation in DTN}

\author{Sobin~CC,~\IEEEmembership{Student~Member,~IEEE}
        Snehanshu~Saha,~\IEEEmembership{Senior~Member,~IEEE}
        Vaskar~Raychoudhury,~\IEEEmembership{Senior~Member,~IEEE}
        Hategekimana Fidele, Sumana~Sinha,~\IEEEmembership{~Member,~IEEE}
        % <-this % stops a space
\thanks{Sobin is a PhD student in the Dept. of Computer Science and Engineering, IIT Roorkee, India. e-mail: sobincc@gmail.com.}% <-this % stops a space

\thanks{Snehanshu is a professor in Dept. of Computer Science and Engineering, PESIT-South Campus, Bangalore, India. e-mail: snehanshusaha@pes.edu.}

\thanks{Vaskar is an Alexander von Humboldt Fellow in Universität Mannheim and Technische Universität Darmstadt and an assistant professor (on leave) in the Dept. of Computer Science and Engineering, IIT Roorkee, India. e-mail: vaskar@ieee.org}

\thanks{Hategekimana is a research scholar in the Dept. of Mathematics, Jain University, India. e-mail: fideleh67@gmail.com}

\thanks{Sumana is a PhD student in the Dept. of Computer Science and Engineering, PESIT South Campus, India. e-mail: sumanasinha@pes.edu.}}
\IEEEtitleabstractindextext{%
\begin{abstract}
Delay Tolerant Networks (DTNs) are sparse mobile networks, which experiences frequent disruptions in connectivity among nodes. Usually, DTN follows store-carry-and forward mechanism for message forwarding, in which a node store and carry the message until it finds an appropriate relay node to forward further in the network. So, The efficiency of DTN routing protocol relies on the intelligent selection of a relay node from a set of encountered nodes. Although there are plenty of DTN routing schemes proposed in the literature based on different strategies of relay selection, there are not many mathematical models proposed to study the behavior of message forwarding in DTN. In this paper, we have proposed a novel epidemic model, called as CISER model, for message propagation in DTN, based on amoebiasis disease propagation in human population. The proposed CISER model is an extension of SIR epidemic model with additional states to represent the resource constrained behavior of  nodes in DTN. Experimental results using both synthetic and real-world traces show that the proposed model improves the routing performance metrics, such as  delivery ratio, overhead ratio and delivery delay compared to SIR model.
\end{abstract}

}

\maketitle

\IEEEdisplaynontitleabstractindextext

% \fi
\IEEEpeerreviewmaketitle

\IEEEraisesectionheading{\section{Introduction}\label{sec:introduction}}

Infectious diseases have caused a large number of mortality in recent years (e.g., includes SARS (2003), swine flu (2009) and MERS CoV (2013), etc.). Mathematical modeling of infectious diseases is used to predict the transmission and the outcome of the diseases, which helps to provide possible countermeasures to reduce the mortality rate or to eradicate the diseases. Amoebiasis is a chronic infectious disease, caused by unicellular micro-organism \textit{Entamoeba histolytica}, which is continuously threatening countless human beings living in unhygienic environment/conditions in developing  countries, especially in Sub-Saharan Africa (SSA). Since amoebiasis is an infectious disease, some of the researchers \cite{AMOEBIASIS} have  modeled the transmission behavior of amoebiasis in human population. Inspired from such  modeling, we have observed that amoebiasis disease modeling can also be applied in modeling the epidemic message forwarding in Delay Tolerant Networks (DTNs).

DTNs are mobile networks in which a complete end-to-end path rarely exists due to high node mobility and frequent disconnection. Since, the connectivity between nodes in DTN is not constantly maintained, a routing protocol is required, which tries to route messages through one or more relay nodes in opportunistic multi-hop manner. Epidemic routing \cite{vahdat2000epidemic} is one of the simplest routing protocol schemes in DTN, which adopts a flooding-based strategy for message forwarding. The basic idea is that a source node having a message to a destination, forwards it to all its neighbors. The neighbor nodes act as relays, floods the message further in the network, so that message eventually delivered to the destination.

\begin{figure}[!t]
\centering
\includegraphics[width=3.6 in]{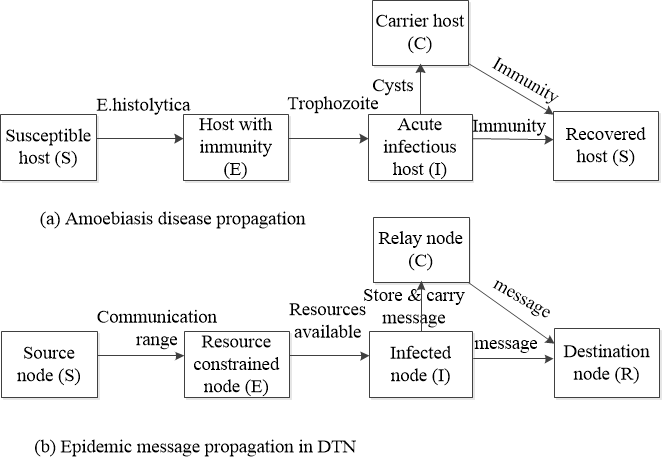}
 \caption{Comparison of (a) amoebiasis disease propagation  with (b) Epidemic message propagation in DTN disease propagation }
\label{fig_ex1}
\end{figure}

There is a clear connection between epidemic message transmission in DTN and the amoebiasis disease propagation in human population (See Fig.\ref{fig_ex1}). Amoebiasis is an infectious disease caused by a parasitic protozoan, Entamoeba histolytica. There are two stages of the life cycle of \textit{Entamoeba histolytica}, infectious cysts and motile phagocyte trophozoites. The infective form of Entamoeba histolytica, called cysts, are shed within the feces of the infected host and later infect food and drinks by flies or other means of direct or indirect contact with contaminated feces. The trophozoites is an acute infectious stage of amoebiasis and exists only in the host. 

A similar scenario exists in case of epidemic message transmission in DTN; when a source node with a message to be delivered to a destination, store and carry the message, until it finds a node in its communication range to forward. The amount of time taken by a node to store the message without forwarding depends on the connectivity of the network. When the source node finds a node in its communication range, it will choose the encountered node as relay and creates a copy of the message and  forward to it.

In order to design a mathematical model for epidemic message transmission in DTN based on the dynamic transmission of amoebiasis, we propose a novel epidemic model, called as CISER model, with two additional classes, \textit{exposed} (\textit{E}) and \textit{carrier} (\textit{C}) to basic SIR model \cite{kermack1927contribution}. The exposed class (\textit{E}) represents the individuals which are exposed to the disease, but the level of infection is not enough to make them infectious.  Similarly, the carrier class (\textit{C}) represents the individuals, which never recover from the disease and spreads it further in the population. While adopting the CISER model to the epidemic message transmission in DTN environment, the susceptible class (\textit{S}) represents the set of nodes in the network, which are available for receiving a copy of the message. The exposed class (\textit{E}) represents the set of nodes which are already having a copy of the message, but are not able to forward the message further in the network due to lack of resources such as, storage and energy. The infected class (\textit{I}) represents the set of nodes which already received a copy of the message and ready to forward that further in the network. The carrier class (\textit{C}), is a sub-category of infected class, which represents   set of relays nodes, which are having a copy of the message and spreads the message further in the network. The co-existence of two classes, I and C, are similar to trophozoites  and  cysts in the transmission of amoebiasis. In the acute infected stage of amoebiasis (\textit{I}), the host is infected, but not spreading the disease to any other hosts; similar to node having a copy of the message, but not forward further in the network. The carrier stage of amoebiasis, occurs when the host remains  acute infection for a period of time (in class I) and later spreads the infection ( class C) by excretion of cysts in their stools. The recovered class (R) represents the nodes which have either received and delivered the message  to the destination or discarded due to expiry of TTL of the message.

In  nutshell, the paper explores the possibilities of introducing 
a novel model for epidemic message propagation in DTN, considering the resource-constrained behavior of DTN nodes. The goal is to improve certain key QoS parameters and to provide emphatic evidence of how the proposed model is better than other models in the same class (basic SIR model). More specifically, we have investigated the certainty of achieving higher delivery ratio, lower delivery delay and lower overhead ratio compared to the basic SIR model.

\section{Related Work}

The initial attempt to formulate a mathematical model for epidemic disease propagation was by Hamer, et.al \cite{hamer1906milroy} using discrete time model. Later, many improvements  have been proposed, which finally led to the most commonly used model, called \textit{Susceptible}, \textit{Infected} and \textit{Recovered} (SIR) model  proposed in 1927 by Kermack, et. al\cite{kermack1927contribution}. In SIR model, a fixed population can be grouped into either of the three classes, susceptible, infected and recovered. The susceptible class, represent the individuals, who are not yet infected by the disease and so might fall prey at any instant. The infected class represents, individuals who are already infected and are capable of infecting the susceptible individuals. The recovered class represents the set of individuals who got infected and then recovered either due to immunization or due to death. Subjects in this class are not infected again or can transmit the infection to others.  So, the model flows as S $\rightarrow$ I $\rightarrow$ R. Later, many elaborations on SIR model have been proposed in literature, like SIS, MSIR, SIR/C, SIER.

1.	SIS model: A variation of SIR model, where an infected individual does not have any long lasting immunity. E.g. diseases like cold and influenza, where upon recovery from infection, an individual move on to susceptible class.

2.	MSIR model: An extension of SIR model, where an individual enters into Maternally-derived immunity (M) class instead of susceptible class. E.g., after birth, babies have inbuilt immunity to many diseases like measles.

3.	SIR/C model: A Carrier (C) class is added to the SIR model to represent individuals who never recover from diseases like tuberculosis and carry the infection although they do not spread the disease. 

4.	SEIR model: A modification of the SIR model with an additional Exposed class (E) to represent the individuals which are infected the disease, but are not infectious because of the immunity towards the disease.  

In a recent survey \cite{sobin2016survey}  on routing in DTN environment, we  have pointed out the need for generalized mathematical framework for message forwarding in DTN as an open problem. Without such a generic model, some of the researchers [6-10] have modeled epidemic routing in DTN using aforementioned models to capture infectious disease propagation pattern in human population. Groenevelt, et. al \cite{groenevelt2005message}, have modeled the expected message delay for Epidemic and 2-hop routing schemes based on number of nodes in the network and the inter-contact time between nodes using Markov chain. The authors have also found a probability distribution function for number of  copies of the message in the network at the time of delivering the message to the destination. Based on the results, the values of inter-contact time for different mobility models like Random Way Point (RWP), Random Direction (RD) and Random Walk (RW), etc., are also obtained.

A similar work carried out by Zhang, et.al \cite{zhang2007performance} in which, a unified framework has been proposed for DTNs based on Ordinary Differential Equation (ODE) to model the message delay, occupancy of buffer space in a node and the number of  copies of the message in the network. Both the schemes \cite{groenevelt2005message}\cite{zhang2007performance} followed the SIR model of infectious disease propagation to model the message delay in DTN. Later, many of the researchers \cite{jacquet2010information}\cite{lin2008stochastic}\cite{hsu2007modeling}\cite{jindal2006performance} modeled message delivery for different DTN applications. However, many of them use only the basic SIR model. To the best of our knowledge, our paper is the first to model epidemic message propagation using an extension of the basic SIR model with two additional states \textit{Exposed} and \textit{Carrier}.

The difference between SIR and CISER model is that, in SIR model, a victim may be infected again irrespective of the immunity and vaccination. However, for many of the diseases, an individual who is having contact with an infected person may not be always  infectious because of his immunity towards the disease. Such class of individuals are not captured in the SIR  model. In CISER model, the exposed class (\textit{E}) represents  population of individuals, who are not infectious, based on their resistance towards the disease. Similarly, for  diseases like amoebiasis, in  acute infectious stage (trophozoites) the infection retains within the host itself. Only in the infectious carrier stage, the disease is spread to individuals by excretion of cysts. However, with basic SIR model, both  types of individuals are combined in only one class of population (I), whereas CISER model considers two different classes \textit{I} and \textit{C} for representing acute infectious stage and infectious carrier stage separately. Therefore, CISER is a more generalized model, which captures more possibilities as compared to SIR model.

\section{Background of the Amoebiasis and its relation with DTN}

\begin{figure}[!t]
\centering
\includegraphics[width=2.6 in]{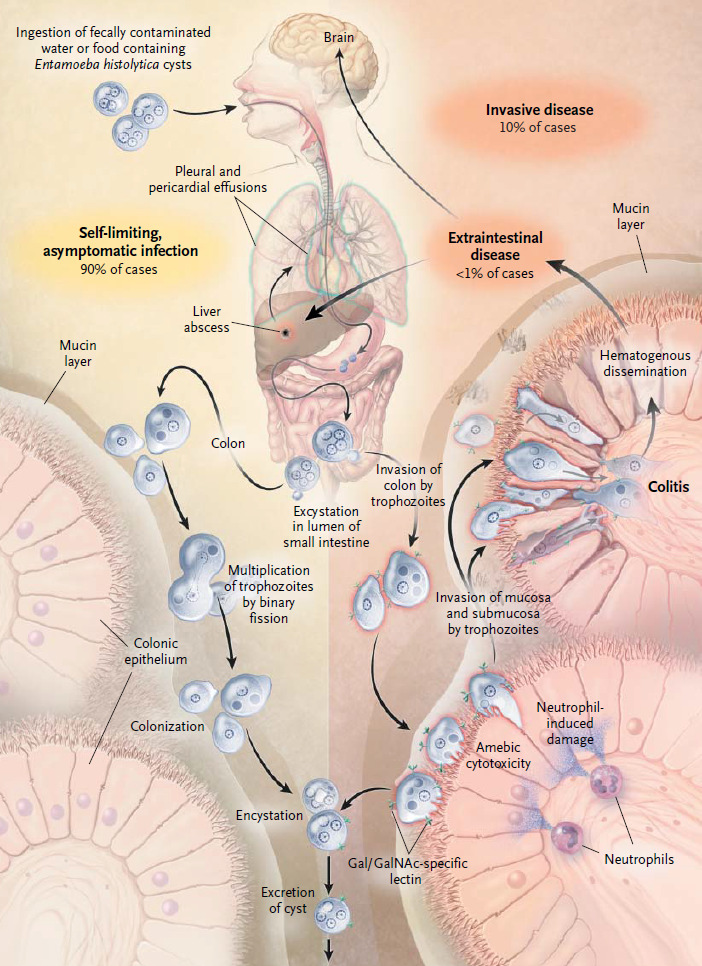}
 \caption{Life cycle of  amoeba \cite{haque2003amebiasis} }
\label{fig_ex}
\end{figure}

Entamoeba  histolytica, the amoebiasis-causing pathogenic specie, which is even able to evade and harm other internal human organs such as the brain, lungs and liver. Entamoeba histolytica is also counted among group of chronic, disabling, and disfiguring diseases; commonly called the Neglected Tropical Diseases (NTDs). They cause naturally degrading effects on the poor and on some disadvantaged urban populations where the conditions of sanitation are very critical. There is a clear interconnection between the life cycle of Entamoeba histolytica and the disease amoebiasis for the last is the end of this cycle. The life cycle of Entamoeba histolytica revolves around two stages of life: infectious cysts and motile phagocyte trophozoites (10 to 60 μm) [2,9]. Entamoeba histolytica in infective forms, called cysts, of radial dimension in the range of 10 to 15μm, are shed within the feces of the infected host and later infect food and drinks by flies or other means  of direct or indirect contact with contaminated feces.

The life cycle of amoeba E. Histolytica involves many microscopic steps of development. Initially cysts are ingested, very soon becomes mature and start a metacystic amoeba stage called trophozoites or sporozoites. These trophozoites go through reproduction by binary fission and cell division. Once the trophozoites arrive in the intestinal tract of the host, they grow and cause the cell invaded to die; this is the beginning of the amoebiasis disease. The trophozoites finally creates cyst, which is incorporated in fecal waste and leave the host in large numbers.  The Fig. \ref{fig_ex} retrieved from the indicated source gives a detailed commented life cycle of amoeba.

The transmission dynamics of amoebiasis (Fig. \ref{fig_ex1} (a)) resembles epidemic message propagation in DTN (Fig. \ref{fig_ex1} (b)).  In DTN, because of frequent network disconnections,  a source node has to store and carry the messages, until it finds a neighbor (relay) node in its communication range to forward. The waiting time for getting  contact opportunity depends on the dynamic network topology, which is similar to the infection stage of amoebiasis. Suppose, if the source node found a relay node and forwards the message, then the relay node may not forward it further in the network because of the scarcity of its resources, such as battery power and storage space. Such  class of relay nodes are similar to  hosts with already infected amoebiasis, but are not infectious (exposed). The relay nodes with sufficient resources and are having a message copy (infected), store the message and carry it, until it finds the destination  or another relay node in its communication range. If such an opportunity exists, infected node will create a copy of the message and forwards the message (carrier node), further in the network.

In next section we will discuss the proposed CISER model for epidemic message propagation based on amoebiasis disease modeling.

%\begin{figure}[!t]
%\centering
%\includegraphics[width=3.5 in]{ameobiasis_DTN2}
% \caption{Epidemic message propagation in DTN}
%\label{fig_ex2}
%\end{figure}

\section{Mathematical Modeling of Epidemic Message Propagation }

The dynamism of CISER model for epidemic message propagation can be represented using a set of nonlinear differential equations based on the Initial Value Problem (IVP), in the following general form.

\begin{equation} \label{eq:0}
\frac{dY}{dt} = f(t,Y,\zeta), Y(t_{0}) =Y_{0}
 \end{equation} 

where Y $\in \textit{IR}^{n} $, Y = Y (t) is a vector, which defines the size of the composition of different classes of nodes and $\zeta$ is a vector for representing the parameters of the CISER model. The DTN network under consideration is grouped into classes of nodes, where at any time \textit{t} of epidemic message propagation, the nodes in the network can be classified into one of the following classes:

$\bullet$ Susceptible \textit{S(t)}: The susceptible class represents the set of nodes in  the network, which are not yet infected, but still susceptible to the infection, i.e, the nodes which are available for receiving the message copies. 

$\bullet$ Exposed \textit{E(t)}: TThe class of  nodes already having a copy of message, however, due to want of resources such as storage space and energy,  message propagation is not yet started.

$\bullet$ Infected \textit{I(t)}: The class of nodes which are having a copy of the message and waiting for a contact opportunity to forward  it further in the network.

$\bullet$ Carrier \textit{C(t)}: The class of  nodes, who host the infection and can spread  the infection further in the network. With respect to epidemic message propagation, the nodes in this class represent the relay nodes, which have enough resources to forward the message further in the network.
 
$\bullet$ Recovered \textit{R(t)}: Class of  nodes which either already received the message or discarded because of expiry of Time To Live (TTL) of the message.

The size of the above classes w.r.t to the proportions of the network can be expressed  as follows.

\vspace{0.2 in}

\textit{S}= $\frac{S(t)}{N}$ , \textit{E}= $\frac{E(t)}{N}$, \textit{I}= $\frac{I(t)}{N}$, \textit{C}= $\frac{C(t)}{N}$ and \textit{R}= $\frac{R(t)}{N}$

\vspace{0.2 in}

where,  \textit{N} is  the  total  number  of  the  nodes in the network.  Let $\frac{1}{\lambda}$, $\frac{1}{\sigma}$, $\frac{1}{\gamma}$, $\frac{1}{\tau}$ and $\frac{1}{\omega}$ be the average periods  of  time  a node  remains  in  Susceptible,  Exposed,  Infected,  Carrier  and  Recovered classes,  respectively. Assume $\lambda$ to  be  infection (contact) rate  at  which  the susceptible nodes acquire the infection.

\vspace{0.2 in}

 $\lambda$ = $\beta \Bigg[ \frac{I(t)}{N} + \epsilon \frac{C(t)}{N} \Bigg] $ or $\lambda = \beta( I + \epsilon C)$
 
 \vspace{0.2 in}

Here $\beta$  is contact rate  among nodes in the network, so that the message can be forwarded to nodes in the transmission range (transition from susceptible to infected). Also, $\epsilon\beta$ indicates the reduction in message transmission due  to  the  noise and carrier  component \cite{keeling2008modeling}.

Assume that the lifetime of a node is  $\frac{1}{\mu}$ (depends on remaining battery life) and  $\rho$ is the probability for an infected node to become a carrier node (based on its resource constraints). Also, (1-$\rho$) is the probability that an infected node become recovered (either destination of the message or message TTL expired). We also assume that no new nodes enter into the network and a small portion of nodes will die because of the exhaustion of  their battery life, which is represented by $\mu$.

\begin{equation} \label{eq:5}
\frac{dS}{dt} =-(\beta I + \epsilon\beta C)S -\mu S +\omega R
 \end{equation} 

\begin{table}[ht]
\caption{Table of parameters}
\centering
\begin{tabular}{|l|c|c|}
\hline
\textbf{Parameter} &\textbf{ Description }  \\
 \hline
$\lambda$ & Average time period of a susceptible node  \\
 \hline
 $\beta$ & Direct transmission rate (contact rate)\\
  \hline
$\epsilon$ & Transmission reducing factor\\ 
\hline 
$\sigma^{-1}$& Average time period of a exposed node\\
\hline 
$\gamma^{-1}$& Average time period of infection\\
\hline 
$\omega^{-1}$&Average  time period of  recovery\\
\hline 
$\tau^{-1}$&Average time period of a carrier  node  \\
\hline 
$\mu$ & Death rate of nodes \\
\hline
$\rho$&Probability that an infected node becomes carrier \\
\hline

\end{tabular}
\label{tab:parameters}
\end{table}

\subsection{Detailed explanation}

The flowchart of the CISER model for epidemic message propagation is given in (Fig.\ref{fig_flowchart}), which is inspired from dynamics transmission of amoebiasis disease propagation \cite{AMOEBIASIS}. The rectangles in the flowchart represent the different classes of the nodes in the network. The transition from one class of nodes to another is represented with the help of arrows. The dead nodes are represented using circles.

The CISER model of epidemic message propagation, proposed for the first time by the authors, is illustrated in the flow chart is explained as follows:

\begin{enumerate}

  \item In the process of epidemic message propagation, a proportion of susceptible nodes in the network in contact with  the infected nodes, receive a copy of the message, and thereby moving to the exposed class with a rate of $\lambda $. A small set of susceptible nodes experience natural death because of drainage of their battery life with a rate  proportional to $\mu$. Also, the number of susceptible nodes is increased from the nodes, which are recovered (discarded the message because of expiring TTL of the message), with a rate of  $\omega$. So, the rate of change in size of susceptible nodes can be represented in terms  of following differential equation using the principle of the law of mass action \cite{segel2013primer}.

\begin{equation} \label{eq:5}
\frac{dS}{dt} =-(\beta I + \epsilon\beta C)S -\mu S +\omega R
 \end{equation}

\item The susceptible portion of nodes infected move to the class of exposed, they will remain in this class for a period of the duration $\frac{1}{\sigma}$. However, during this period of stay, there are two possible issues with this proportion of the exposed individuals in this class \textit{E}: either they become infected at a rate proportional to $\sigma$ or some of them may experience natural death at a rate proportional to $\mu$. Mathematically, the net change of the total exposed hosts in this population at any time during the course of message propagation, is denoted by $\frac{dE}{dt}$. It follows that the equation expressing the rate of change in size of the exposed proportion of the population is

\begin{equation} \label{eq:6}
\frac{dE}{dt} =(\beta I + \epsilon\beta C)S -(\sigma + \mu) E
 \end{equation}

\item In the middle age of the dynamics of epidemic message propagation, there are two categories of the nodes who are already infected namely \textit{I} and \textit{C}. Most of the time, when the proportion of exposed population becomes infected, their infected state may be either acute or latent and they are able to spread out the infection in proportions of respective probabilities $\beta$ and $ \epsilon\beta$, where 0 $< \epsilon <$ 1. We assume that the coexistence of these two states in approximate proportions of 20$\%$ and 80$\%$ respectively, and for this reason, the model suggests there is a probability $\rho$ that an infected node is being a carrier. During the period of infection, $\frac{1}{\lambda}$,  the acute infected nodes  leave this stage at rates proportional to $\gamma$ but taking account of the probability $\rho$, i.e., $\rho \gamma$ to become carriers while the remaining proportion $(1 - \rho ) \gamma$ recovers and enters the class \textit{R}. Note that infected carriers will not remain forever, they will decrease as they recover at the rate  $\tau$. Also, the death of infected nodes will also decrease the size of the populations in both compartments \textit{I} and \textit{C} at the rate equivalent to $\mu$. Mathematically, the dynamics of message propagation over these two classes is expressed in term of the following ordinary differential equations:

\begin{equation} \label{eq:7}
\frac{dI}{dt} =\sigma E -(\gamma + \mu) I
 \end{equation} 
 
\begin{equation} \label{eq:8}
\frac{dC}{dt} =\rho \gamma I -(\tau + \mu) C
 \end{equation}  
 
  \begin{figure}[!t]
\centering
\includegraphics[width=3.5 in ]{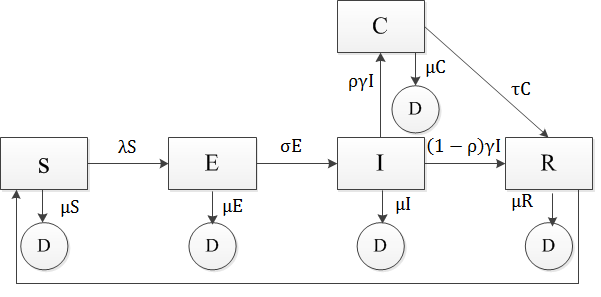}
% where an .eps filename suffix will be assumed under latex, 
% and a .pdf suffix will be assumed for pdflatex; or what has been declared
% via \DeclareGraphicsExtensions.
\caption{Flow chart of the system dynamics}
\label{fig_flowchart}
\end{figure}

\item As far as the dynamics of epidemic message propagation on the recovered class \textit{R} is concerned, the ordinary differential equation \ref{eq:9} sums up different transfers taking place as the infected nodes have cleared the infection either by delivering the message (to another node/destination) or discarded the message (if its TTL is expired). This equation summarizes how the size of \textit{R} change continuously during the period of message propagation. During the recovery period of length $\frac{1}{\omega}$, a  fraction of population proportional to the size of \textit{R} is removed from the proportion of the population admitted in the recovered class $(1 - \rho ) \gamma I + \tau C $, at the rate $\omega$ and then returned to the susceptible class. At the same moment, some of the node's battery life may exhaust and thereby suffer natural death and yield an additional proportional of hosts to be removed from the class of recovery \textit{R} proportionally to its size at a rate of $\mu$.

\begin{equation} \label{eq:9}
\frac{dR}{dt} =(1-\rho) \gamma I +\tau C -(\omega + \mu)R
 \end{equation}  
 
\end{enumerate}

These five ordinary differential equations (Eq. \ref{eq:5} to \ref{eq:9}) coupled with the flowchart (Fig.\ref{fig_flowchart}) form a system that governs the dynamics of epidemic message propagation through the population. A solution to this system is a vector function that provides, at any time \textit{t} of the course of epidemic message propagation, the coordinates of the point in five dimensional space whose components are expressed in terms of sizes of susceptible, exposed, infected, carrier and recovered respectively. From the system of ordinary differential equations, it follows that

\begin{equation} \label{eq:10}
\frac{dS}{dt} + \frac{dE}{dt} + \frac{dI}{dt} + \frac{dC}{dt} + \frac{dR}{dt}=0
 \end{equation}

 Integrating Eq.\ref{eq:10} with respect to the time, the integral yield the following result
 
 \begin{equation} \label{eq:11}
 S(t) + E(t) + I(t) + C(t) + R(t) =k
 \end{equation}  
 
where \textit{k} is the constant of integration which is equal to 1 at the initial time \textit{t} = 0, In other words, we should say that at any time of the outbreak of epidemic message propagation, the total proportions of the  susceptible,  exposed,  infectious,  carrier  and  removed  is  equal  to  1.  Epidemiologically, we  can conclude that the size of the population under consideration does not change during the whole period of the course of epidemic message propagation. 

\begin{equation} \label{eq:12}
 S(t) + E(t) + I(t) + C(t) + R(t) =1, \hspace{0.3in} \forall t \geq t_{0}
 \end{equation}

As  the  consequence  of  the  Eq.\ref{eq:12},  it  is  clear  that  one  variable  of  the  system  of  ordinary  differential  equations    can  be  expressed  in  terms  of  the  remaining  others. To  avoid  the  redundant equations,  the  Initial Value Problem  (IVP) coupled with  the  flowchart (Fig.2) and describing  the nature and characteristic of the dynamics of message propagation in a given population can be expressed as follows.

\begin{equation} \label{eq:15}
\frac{dS}{dt} =-(\beta I + \epsilon\beta C) -\mu S +\omega R
 \end{equation}

\begin{equation} \label{eq:16}
\frac{dE}{dt} =(\beta I + \epsilon\beta C)S -(\sigma + \mu) E
 \end{equation} 

\begin{equation} \label{eq:17}
\frac{dI}{dt} =\sigma E -(\gamma + \mu) I
 \end{equation} 
 
\begin{equation} \label{eq:18}
\frac{dC}{dt} =\rho \gamma I -(\tau + \mu) C
 \end{equation}  
 
Subject to the initial condition: 

\vspace{0.2 in}

S(t$_{0}$)= S$_{0} \geq 0$, E(t$_{0}$)= E$_{0} \geq 0$, I(t$_{0}$)= I$_{0} \geq 0$, C(t$_{0}$)= C$_{0} \geq 0$
\vspace{0.2 in}

The vector  \textit{x} introduced in the general form of the initial value problem (Eq. \ref{eq:0}) is characterized by  its  components  which  are  the  proportions  of  the  individuals  in  each  class,  i.e.,
x(t) = (S, E, I, C) $\in \Omega$,    with $ \Omega \subset IR^{4}$ the  domain  of  the  IVP,  here  called  the  phase  space,  is defined  as  $\Omega$ = {(S, E, I, C) : S + E + I + C $\leq$ 1, S, E, I , C $\in IR^{+}$ } which  is  positively  invariant under the vector field defined by (Eq. \ref{eq:0}). For  this reason, any trajectory starting  in  the phase space  $\Omega$ remains inside the domain for all time t $ \geq t_{0}$ \cite{hethcote2000mathematics}. In addition to the positive invariant of the domain, the  partial  derivatives  of  the  vector  function  in  the  right  side  of  (Eq. \ref{eq:0})  with  respect  to  the  second variable  \textit{x} as well as  to $\eta$  are continuous. For  this  reason,  \textit{f (t, x; $\eta$)} is  smooth on  its domain.  In view of  these  facts, unique solutions exist  in  $\Omega$ for all  t $\geq$ 0  \cite{wiggins2003introduction} and  the model  is mathematically and epidemiologically well posed. 

\subsection{Dead population prediction}

Let \textit{D(t)} denote  the  number  of  the  nodes  dying in the process of message propagation by exhausting their battery power. When message propagation begins, the host  population  is  supposed  to  suffer  the  natural  death.  Referring  to  the  flowchart,  instantaneous variation of \textit{D(t)} is given by:

\begin{center}
$\frac{dD}{dt} = \mu( S(t) + E(t) + I(t) + C(t) + R(t))$
\end{center}

\begin{center}
$dD(t) = \mu dt$
\end{center}

\begin{equation} \label{eq:18_a}
D(t) = D_{0} + \mu t, \hspace{0.1 in} D_{0}= D(0)
 \end{equation}  
 
Note that the value of $\mu$ is very small and depends on the duration of message propagation.

\subsection{Qualitative study of the model}

The most  important  in  the  dynamics  of epidemic message propagation,  is  to  quantify  the  ability   to spread out the message through the population.  In other words,  the possibility of the message propagation  to  invade the population depends on the value of  threshold parameters.  In  the  literature of infectious diseases, the basic reproduction number \textit{R$_{0}$} is the determinant parameter for the spread of the disease. The  basic  reproduction  number  is  defined  as  the  number  of  people  infected  by  only  one  typical infection  introduced  into  the  entire  population  of  susceptible  individuals  \cite{keeling2008modeling}\cite{anderson1991infectious}\cite{diekmann1990definition}\cite{van2002reproduction}. 

As far as the threshold  conditions of message propagation is concerned,  to  spread  out,  the  basic  reproduction  number must  be greater than 1, otherwise message propagation will die off, i.e.,  \textit{R$_{0}$} $>$ 1.  The message propagation starts only  if  one  such  infected  node  can  pass  on  the  message  on  average  to  at  least  one susceptible.  The  basic  reproduction  number  \textit{R$_{0}$} of  this  model  is  determined  by  using  the  \textit{Next Generation Operator} (NGO), a method derived from the theory of center manifold and developed by Driessche, et.al\cite{van2002reproduction} and will be explained in the following subsection.

\subsubsection{Determination of the basic reproduction}

Let  \textit{X} be the  column  vector  of  the  infected  compartments  i.e.,  \textit{X} = [E, I, C]$^{\prime}$ and  let  \textit{Y} denotes  the column  vector  of  the  remaining  classes,  that  is  \textit{Y} = [S, R]$^{\prime}$   and we rearrange  the  Equ.[\ref{eq:5}-\ref{eq:9}] in such a way that we obtain an equivalent system of the following form: 

\begin{equation} \label{eq:13}
\frac{dX}{dt} = \Phi(X,Y) - \Psi (X,Y)
 \end{equation}

 \begin{equation} \label{eq:14}
\frac{dY}{dt} = W (X,Y)
 \end{equation}

Where $\Phi$(X,Y) is  the vector  function of  the new  infection  rates  or  function expressing  the  flow from uninfected class \textit{Y} to the classes of infected classes \textit{X} and  $\Psi$(X,Y) is a vector function of all others rates entering the infected classes in Eq.\ref{eq:13}, affected with a negative sign. Let us denote the \textit{Message Propagation  Equilibrium} (\textit{MPE}) point by P$_{0}$= [$\overline{X}, \overline{Y}$], where $\overline{X}$ =0, i,e. P$_{0}$= [0, $\overline{Y}$]. The vector functions $\Phi$ and $\Psi$  must satisfy  $\Phi(P_{0}$) = 0  and  $\Psi(P_{0}$) = 0. Since,  the MPE occurs when $\frac{dx}{dt}$= 0, combined with  the condition of  total extinction of  the disease, i.e.,  \textit{E} = 0,  \textit{I} = 0  and  \textit{C}= 0. Solving  the  system of  equations  obtained  by  introducing  these  two conditions in Eq.[\ref{eq:5}-\ref{eq:8}], we obtain the MPE P$_{0}$= [0, 0, 0,1, 0]  and hence $\overline{Y}$ = [1, 0]. Now, define two matrices  \textit{F}  and \textit{V}  by  

 \begin{equation} \label{eq:14_a}
F=  \frac{\partial \Phi}{\partial X} \vert_{P_{0}} 
 \end{equation}
 
 \begin{equation} \label{eq:14_a}
 V=  \frac{\partial \Psi}{\partial X} \vert_{P_{0}} 
 \end{equation}

The Next Generation Operator method  says  that  the  basic  reproduction  number, R$_{0}$, is  equal  to  the spectral  radius of the Next Generation Matrix,  FV$^{-1}$. Apply  the above to the dynamics of message propagation, we deduce from the sub-system of equations \ref{eq:13} expressions for the vector functions  F(X,Y)  and  Y(X,Y) considering its equivalent system defined by the equations \ref{eq:6}, 
\ref{eq:7} and \ref{eq:8}, i.e., the system of differential equations 

\begin{equation} \label{eq:19}
\frac{dE}{dt} =(\beta I + \epsilon\beta C) S -(\sigma + \mu) E
 \end{equation} 

\begin{equation} \label{eq:20}
\frac{dI}{dt} =\sigma E -(\gamma + \mu) I
 \end{equation} 
 
\begin{equation} \label{eq:21}
\frac{dC}{dt} =\rho \gamma I -(\tau + \mu) C
 \end{equation}

It follows that

\[  \Phi(X,Y) = \left| \begin{array}{c}
(\beta I + \epsilon\beta C)S  \\
0 \\
0  \end{array} \right|\]

and

\[  \Psi(X,Y) = \left| \begin{array}{c}
(\sigma + \mu) E  \\
-\sigma E +(\gamma + \mu) I \\
-\rho\gamma I +(\tau + \mu) C  \end{array} \right|\]

To find the matrices F and V

\[ F=  \frac{\partial \Phi}{\partial X} \vert_{P_{0}}   = \left| \begin{array}{ccc}
0 & \beta S & \epsilon\beta S  \\
0 &0 & 0 \\
0 &0 & 0 \end{array} \right|\]

with initial condition ${P_{0}} $ =[0, 0, 0, 1]

 \[ F =  \left| \begin{array}{ccc}
0 & \beta  & \epsilon\beta   \\
0 &0 & 0 \\
0 &0 & 0 \end{array} \right|\]

\[ V=  \frac{\partial \Psi}{\partial X} \vert_{P_{0}}   = \left| \begin{array}{ccc}
(\sigma +\mu) & 0& 0  \\
-\sigma &(\gamma+ \mu) & 0 \\
0 &-\rho\gamma & (\tau+\mu) \end{array} \right|\]

V$^{-1}$= 

\[   \frac{1}{\left|V\right|}\left| \begin{array}{ccc}
(\gamma +\mu)(\tau+\mu) & 0& 0  \\
\sigma(\tau+\mu) &(\sigma+\mu)(\tau+\mu) & 0 \\
\sigma\rho\gamma & \rho\gamma(\sigma+\mu) & (\sigma+\mu)(\gamma+\mu) \end{array} \right|\]

with $\left|V\right|$ = $(\sigma+\mu) (\gamma+\mu)(\tau+\mu)$

\vspace{0.3 in}

FV$^{-1}$=
\[    \frac{1}{\left|V\right|}\left| \begin{array}{ccc}
a & b & c  \\
0 &0 & 0 \\
0 & 0 & 0 \end{array} \right|\]

where, a= $\beta\sigma(\tau+\mu)+\epsilon\beta\sigma\rho\gamma$, b= $\beta(\sigma+\mu)[(\tau+\mu)+\epsilon\rho\gamma]$ and c= $\epsilon\beta (\sigma+\mu)(\gamma+\mu) $. The  spectral  radius  of  the  Next  Generation Matrix, $FV^{-1}$ which  gives  the  value  of  the  reproduction number as:

\begin{equation} \label{eq:21_2}
R_{0} = \frac{\beta\sigma(\tau+\mu)+\epsilon\beta\sigma\rho\gamma}{(\sigma+\mu)(\gamma+\mu)(\tau+\mu)} 
 \end{equation}

or

\begin{equation} \label{eq:22}
R_{0} =\frac{\beta\sigma}{(\sigma+\mu)(\gamma+\mu)}\Big[ 1 + \frac{\epsilon\rho\gamma}{(\tau+\mu)}\Big]
 \end{equation}

In  accordance  with  the meaning of  the  basic  reproduction  number  and  its  properties,  the  outbreak of message propagation is expected to occur at the end of the latent period,  if the set $\eta$  of parameters satisfies the conditions,

\begin{equation} \label{eq:22_1}
 R_{0} = \frac{\beta\sigma}{(\sigma+\mu)(\gamma+\mu)}\Big[ 1 + \frac{\epsilon\rho\gamma}{(\tau+\mu)}\Big] > 1
 \end{equation}

 Otherwise, the  message propagation  will  never  spread  out  even though  there  should  be  a  negligible  number  of  infected  nodes  in  the  population.  In the dynamics of epidemic message propagation,  any mechanism of monitoring and prevention of this message propagation will focus on the effects resulting from variations of the basic reproduction number.
 
 In  view of  Eq. \ref{eq:22},  the reproduction  number  tends  to  be zero  as  the  transmission  is  very  negligible,  we  express  this  fact mathematically by  R$_{0} \rightarrow$ 0  as $\beta \rightarrow$ 0. This limit occurs only if either the adequate contacts are very limited or the probability of transmission of the message, when  in  contact with infectious node is nearly zero. Under such conditions, the basic reproduction number will be kept less than 1 and then the dynamics of message propagation will settle in the  equilibrium state.

\subsubsection{Endemic equilibrium}

The endemic equilibrium occurs when  the  state of  the  system does not vary over  time. The coordinates of the endemic equilibrium  are denoted by ($\overline{S}, \overline{E}, \overline{I}, \overline{C}, \overline{R})$

\begin{equation} \label{eq:23}
\frac{dx}{dt} =0, \overline{I} =c >0, c\in IR
 \end{equation} 

With Eq. \ref{eq:23}, we are expecting a solution in the form \\ x(t) = k  \cite{perko2001nonlinear}\cite{wiggins2003introduction}\cite{segel2013primer}, where k= ($\overline{S}, \overline{E}, \overline{I}, \overline{C}, \overline{R})$. In other  words,  the  numbers  of  individual  in  each  compartment  remain  the  same  during  endemic equilibrium of message propagation or the rate of individual entering and exiting each class are equal. In other words the endemic equilibrium is the state at which message propagation persists as long as this state is stable.

\begin{theorem}
%\emph{(Endemic equilibrium)}
\label{Endemic}
At endemic equilibrium of message propagation, the product of the proportion of susceptible  individuals and the basic reproduction number is equal to 1.
\end{theorem}

%\newenvironment{proof}[1][Proof]{\begin{trivlist}
%\item[\hskip \labelsep {\bfseries #1}]}{\end{trivlist}}

%\begin{proof}

\textit{Proof:}
Rewriting the Eq. \ref{eq:23} into its components yields the following nonlinear system of equations: 
%\end{proof}

\begin{equation} \label{eq:23_1}
\mu- (\beta \overline{I} +\epsilon\beta\overline{ C})\overline{ S} -\mu \overline{ S} + \omega \overline{ R}=0
 \end{equation} 

\begin{equation} \label{eq:24}
(\beta \overline{I} +\epsilon\beta\overline{ C})\overline{ S} -(\sigma +\mu) \overline{ E}=0 
 \end{equation} 

\begin{equation} \label{eq:25}
\sigma \overline{ E}  -(\gamma+\mu) \overline{ I} =0
 \end{equation}

\begin{equation} \label{eq:26}
\rho \gamma \overline{ I}  -(\tau+\mu) \overline{ C} =0
 \end{equation}

\hspace{1.2 in}

Departing to the Eq. \ref{eq:25} and \ref{eq:26}, express  $\overline{ E}$ and  $\overline{ C}$ in terms of  $\overline{I}$

\begin{equation} \label{eq:27}
\overline{ E} =\frac{\gamma+\mu}{\sigma} \overline{ I} 
 \end{equation}

\begin{equation} \label{eq:28}
\overline{ C} =\frac{\rho\gamma}{\tau +\mu} \overline{ I} 
 \end{equation} 

Substituting equations \ref{eq:27} and \ref{eq:28}, in \ref{eq:24}, we obtain

\begin{center}

 $\Big(\beta \overline{I} + \epsilon\beta \frac{\rho\gamma}{\tau+\mu}\overline{I}  \Big) \overline{S} -(\sigma +\mu)\frac{(\gamma+ \mu)}{\sigma} \overline{I} $ =0
\end{center} 

\begin{center}
$ \Big[  \Big(\beta  + \epsilon\beta \frac{\rho\gamma}{\tau+\mu} \Big) \overline{S} -\frac{(\sigma +\mu)(\gamma+ \mu)}{\sigma}  \Big] \overline{I}$ =0
 
 \end{center}   

  Since, $\overline{I} \neq 0$, that is
  
  \begin{center}
$  \Big(\beta  + \epsilon\beta \frac{\rho\gamma}{\tau+\mu} \Big) \overline{S} -\frac{(\sigma +\mu)(\gamma+ \mu)}{\sigma}$ =0
  \end{center}

   \begin{center}
$  \Big(    \frac{(\beta(\tau+\mu)+\epsilon\beta\rho\gamma}{\tau+\mu} \Big) \overline{S} = \frac{(\sigma +\mu)(\gamma+ \mu)}{\sigma} $ 
  \end{center}

   \begin{center}
$ \overline{S} = \frac{(\sigma +\mu)(\gamma+ \mu)(\tau+\mu)}{\sigma  \Big( (\beta(\tau+\mu)+\epsilon\beta\rho\gamma \Big)} $ 
  \end{center}

    \begin{center}
$ \overline{S} = \frac{1}{R_{0}} $ 
  \end{center}

    \begin{center}
$ \overline{S}R_{0} = 1 $ 
  \end{center}

  This theorem  enlightens possible scenarios that may take place when message propagation dynamics is at its  endemic  steady  state.  The  basic  reproduction  number  being  equal  to  the  reciprocal  size  of  the susceptible proportion in this state, by a glimpse of the reproduction number, in the neighborhood of the points of  \textit{S}, we should predict the following behavior of message propagation in terms of spread:

\begin{enumerate}

\item \textbf{Extinction of susceptible population:} Mathematically,  this  situation  happens  as  long  as  $\overline{S}$   tends  to  be zero  (i.e.,  $\overline{S} \rightarrow 0$ )  for  that $R_{0} \rightarrow \infty$.  If  the  reproduction  of  is  very  high,  high  number  of  nodes will  infect  by  one infected node and an overflow of infected nodes will take place. This case occurs when the rates $\lambda ,\alpha$ are very high compared  to  remaining  parameters.  In  this case,  the  time  taken by an  infected node  to  recover will be excessively  long. The whole population will completely spread the message forever and ever, i.e., a node will receive duplicate messages again and again without termination.

\item \textbf{Eradication of message propagation:} At endemic equilibrium or steady state, if by intervention method, the period of recovery $\gamma$  and  that  of  the  decay  of  immunity $\omega$   is  shortened,  a  considerable  number  of  nodes will become susceptible after a short period of message propagation. This means that $R_{0} \rightarrow 0 $, as $\overline{S} \rightarrow 1$. In this case, the size of the nodes in infected classes will drastically decrease to zero and the capability to infect will  drop. The system will revolve toward the message free equilibrium  state. In other words, message propagation will die out.

\item \textbf{Marginal size of the population for message spread:} The theorem \ref{Endemic}, testifies the existence of the  size of  the population under which the message  can  never be able to spread. Under  any  other  circumstances,  onset  of  message propagation would  be impossible  if  the  initial proportion of  susceptible  individuals  $ S_{0} = \overline{S}$   satisfies  the condition $\frac{1}{S_{0}} \leq 1$.  In  practice,  this  condition  holds  if  every  node  in  the  population  is  susceptible  to message propagation.

\end{enumerate}

\begin{theorem}
%\emph{(Endemic equilibrium)}
\label{Endemic1}
The endemic equilibrium of message propagation occurs only if the basic reproduction number is greater than 1.
\end{theorem}

\textit{Proof:}
The theorem \ref{Endemic} is proven as we come to show that all other coordinate components rather than $\overline{S}$ of the endemic equilibrium contain in their algebraic expression the factor $\Big(1-\frac{1}{R_{0}} \Big)$. Substitute $\overline{E}$ and $\overline{C}$  by their equivalent expressed in the RHS of  equations \ref{eq:27}, \ref{eq:28} respectively in  equation \ref{eq:23_1}. 

Consider the fact that $\overline{R}$ = 1-($\overline{S}$+ $\overline{E}$+ $\overline{I}$ + $\overline{C}$), we obtain:

\begin{center}

$ \mu - (\beta I + \epsilon\beta C) \overline{S} - \mu\overline{S} + \omega\overline{R}$  =0

\end{center}

\begin{equation} \label{eq:28_1}
\begin{split}
 \mu - \Big(\beta I + \epsilon\beta \frac{\rho\mu}{\tau+\mu)} \overline{I}\Big) \frac{1}{R_{0}} - \frac{\mu}{R_{0}} + \\ \omega \Big[ 1-\frac{1}{R_{0}}- \frac{\gamma+\mu}{\sigma)} \overline{I}- \overline{I} - \frac{\rho\gamma}{(\tau + \mu)} \Big]  =0 
 \end{split}
 \end{equation}
 
Using algebraic operations the equation \ref{eq:28_1} is reduced to following form:

\begin{equation} \label{eq:28_2}
 \frac{D}{\sigma(\sigma+\mu)}\overline{I}  = (\omega+\mu)\Big(1-\frac{1}{R_{0}} \Big)
\end{equation}

D is a constant defined as

\begin{center}
$ D= (\sigma+\mu)(\gamma+\mu)(\tau+\mu)+ \omega(\gamma+\mu)(\tau+\mu)+ \omega \sigma(\tau+\mu)+ \omega \sigma \rho\gamma$
\end{center}

\begin{equation} \label{eq:30}
\overline{I}= \frac{\sigma}{D}(\sigma+\mu)(\omega+\mu) \Big( 1-\frac{1}{R_{0}}\Big)
 \end{equation}
 
$\overline{I}$ exists because D $\neq$ 0. Other coordinate values follow the equations \ref{eq:27}, \ref{eq:28} and \ref{eq:12}. Hence,

\begin{center}

$ \overline{E}  = \frac{(\gamma+\mu)}{\sigma} \times \frac{\sigma}{D} (\sigma+\mu) (\omega+\mu) (\Big( 1-\frac{1}{R_{0}}\Big) $

\end{center}

\begin{equation} \label{eq:31}
 \overline{E}= \frac{1}{D} (\gamma+\mu)(\sigma+\mu)(\omega+\mu) \Big( 1-\frac{1}{R_{0}}\Big)
 \end{equation}
 
 Also,
 
 \begin{center}

$ \overline{C}  = \frac{\rho \gamma}{(\tau + \mu)} \times \frac{\sigma}{D} (\sigma+\mu) (\omega+\mu) (\Big( 1-\frac{1}{R_{0}}\Big) $

\end{center}

\begin{equation} \label{eq:32}
 \overline{C}= \frac{\rho \gamma \sigma}{D (\tau + \mu)} (\sigma+\mu)(\omega+\mu) \Big( 1-\frac{1}{R_{0}}\Big)
 \end{equation}

\begin{equation} \label{eq:33}
 \overline{R}= 1-( \overline{S}+  \overline{E} +  \overline{I} +  \overline{C})
 \end{equation}
 
 But we know that the domain $\Omega$ is positive invariant and $\Big( \overline{S}+  \overline{E} +  \overline{I} +  \overline{C}\Big) \in \Omega$. The property of belonging to the domain then implies $\Big( 1-\frac{1}{R_{0}}\Big) > $ 0 and hence R$_{0} >$ 1.

 \subsection{Stability analysis of the steady states}
\subsubsection{The stability of the message propagation equilibrium}
\begin{theorem}
The message propagation equilibrium is asymptotically stable if the basic reproduction number is less than one and it is unstable if it is greater than one.
\end{theorem}

\textbf{Proof}:
The message propagation  equilibrium is asymptotically stable if the real part of all eigenvalues of the matrix $A=\left.\frac{\partial f(t,x)}{\partial x}\right\vert_{x=E_0}$ are negative. Otherwise, message propagation  equilibrium is unstable.
It is very easy to realize that 
$$
A= \left[
\begin{array}{cccc}
-\left(\omega+\mu\right)& -\omega&-\left(\beta+\omega\right)&-\left(\epsilon\beta+\omega\right)\\
0&-\left(\sigma+\mu\right)&\beta&\epsilon\beta\\
0&\sigma&-\left(\gamma+\mu\right)&0\\
0&0&\rho\gamma&-\left(\tau+\mu\right)
\end{array}
\right]
$$
The characteristic polynomial of the matrix $A$ is then given by:

\begin{equation}\label{eq:3_4}
\begin{split}
\left\vert A-\lambda I\right\vert=\left(c_4+\lambda\right)[\lambda ^3+(c_1+c_2+c_3)\lambda^2+\\
(c_1c_2+c_1c_3+c_2c_3-\beta\sigma)\lambda+c_1c_2c_3(1-R_0)]
\end{split}
\end{equation}

where $c_1=\sigma+\mu$, $c_2=\gamma+\mu$, $c_3=\tau+\mu$ and $c_4=\omega+\mu$ are all positive. It is clear that one root of the characteristic polynomial is negative $\lambda_1=-c_4$. Other roots are zeros of the cubic polynomial

\begin{equation}\label{eq:3_5}
\begin{split}
P(\lambda)=\left(c_4+\lambda\right)\lambda ^3+(c_1+c_2+c_3)\lambda^2+\\
(c_1c_2+c_1c_3+c_2c_3-\beta\sigma)\lambda+c_1c_2c_3(1-R_0)
\end{split}
\end{equation}

For matter of signs of the real parts of the roots of the polynomial (Eq.\ref{eq:3_5}), we shall make use of Routh-Hurwitz test [30]. According to this test, the real parts of the roots of the cubic polynomial $P(\lambda)=\lambda ^3+a_1\lambda ^2+a_2\lambda+a_3$ are negative if the coefficients of this polynomial satisfy the following four conditions:

\begin{enumerate}
\item $a_1=c_1+c_2+c_3>0$
\item $a_2=c_1c_2+c_1c_3+c_2c_3-\sigma\beta>0$
\item $a_3=c_1c_2c_3\left(1-R_0\right)>0$
\item $a_1a_2-a_3>0$
\end{enumerate}

By the definition of the coefficients of \textit{c}, the first condition holds true. For seek of the conditions that make the other remaining three, consider the following lemma.\\\\
\textbf{Lemma:}\\
If the basic reproduction number is less than one and the condition $(3)$ is true, then the conditions $(2)$ and $(4)$ are also true.\\\\
\textbf{Proof:}\\
\begin{enumerate}
\item Given $R_0<1$,\\\\ 
It follows that $\left(1-R_0\right)>0$.\\\\ 
Hence, $a_3>0$.\\\\ 
This proves the condition $(3)$.\\

\item To prove that if $(3)$ is true, then condition $(2)$ is true.\\

Consider $a_2=c_1c_2+c_1c_3+c_2c_3-\sigma\beta$ \\\\
From equation \ref{eq:22},  $\displaystyle{\sigma\beta=\frac{c_1c_2c_3R_0}{c_3+\epsilon\rho\gamma}}$\\\\

$\displaystyle{a_2=c_1c_2+c_1c_3+c_2c_3-c_1c_2\frac{c_3R_0}{c_3+\epsilon\rho\gamma}}$\\\\
$\displaystyle{a_2=c_1c_2+c_1c_3+c_2c_3-c_1c_2\frac{R_0}{1+\frac{\epsilon\rho\gamma}{c_3}}}$\\\\
$\displaystyle{a_2=c_1c_2\left(1-\frac{R_0}{1+\frac{\epsilon\rho\gamma}{c_3}}\right)+c_1c_3+c_2c_3}$\\\\

Since, $1-\left(1-\frac{R_0}{1+\frac{\epsilon\rho\gamma}{c_3}}\right)>0$ for $R_0<1$,\\\\ It follows that  $a_2>0$\\\\
Hence the proof of the implication.\\\\

\item To show that if $a_2>0$, then $a_1a_2-a_3>0$ \\\\
Given,
$a_1a_2=\left(c_1+c_2+c_3\right)\left(c_1c_2+c_1c_3+c_2c_3-\sigma\beta\right)$\\\\
$a_1a_2=\left(c_1+c_2+c_3\right)\left[c_1c_2\left(1-\frac{R_0}{1+\frac{\epsilon\rho\gamma}{c_3}}\right)+c_1c_3+c_2c_3\right]$\\\\
But,\\\\ $1-\frac{R_0}{1+\frac{\epsilon\rho\gamma}{c_3}}>1-R_0$\\\\
$a_1a_2>\left(c_1+c_2+c_3\right)\left(c_1c_2\left(1-R_0\right)+c_1c_3+c_2c_3\right)$\\\\

Develop the right side of the inequality by applying distributive and considering the second term yield\\

$a_1a_2>c_1c_2c_3\left(1-R_0\right)=a_3$\\

Hence, $a_1a_2-a_3>0$\\

From the lemma, it is very clear that the theorem is proved and the characteristic polynomial (Eq.\ref{eq:3_4}) has four roots whose real parts are all negative. The fact makes the message propagation  equilibrium to be asymptotically stable.
\end{enumerate}

\subsubsection{The stability analysis of the endemic equilibrium point}

The endemic equilibrium point, is the point at which the message propagation, in its course, remains forever. To understand the behavior of epidemic message propagation at this state, it is necessary to know in which manner the message propagation reaches this state and what are the necessities in favor of this state.

In this study, we suggest to investigate the stability of the endemic equilibrium point   by Liapunov method provided by [30]. For the application of this method, it is necessary to transform the IVP, (Eq. \ref{eq:5} to \ref{eq:8}) to  with initial conditions (S(t$_{0}$)= S$_{0} \geq 0$, E(t$_{0}$)= E$_{0} \geq 0$, I(t$_{0}$)= I$_{0} \geq 0$, C(t$_{0}$)= C$_{0} \geq 0$)  into homogeneous one through the transformation defined by:

\begin{equation} \label{eq:44}
x= \overline{x} + y , x \in \Omega, y \in IR^{4}
 \end{equation}

In components,  the above equation can be expressed by  $ S= \overline{S} +y_{1}; E= \overline{E} +y_{2}; I=\overline{I} +y_{3}; C= \overline{C} +y_{4};$ and it results from the differentiation the following identities

\begin{equation} \label{eq:45}
\frac{dS}{dt} =\frac{dy_{1}}{dt}, \frac{dE}{dt} =\frac{dy_{2}}{dt}, \frac{dI}{dt} =\frac{dy_{3}}{dt}, and \frac{dC}{dt} =\frac{dy_{4}}{dt}, 
 \end{equation}

  Using the transformation (Eq.\ref{eq:44}) and the identity (Eq.\ref{eq:45}), the above IVP is reduced to the following system of ordinary differential equations:
  
  \begin{equation} \label{eq:46}
  \begin{split}
\frac{dy_{1}}{dt} = -(\beta y_{3} + \epsilon \beta y_{4})\overline{S} - (\beta y_{3} + \epsilon \beta y_{4}) y_{1} - (\beta \overline{I} \\ + \epsilon \beta \overline{C} -(\omega+ \mu)) y_{1} - w(y_{2}+ y_{3} + y_{4})
\end{split}
 \end{equation}

   \begin{equation} \label{eq:47}
     \begin{split}
\frac{dy_{2}}{dt} = (\beta y_{3} + \epsilon \beta y_{4})\overline{S} + (\beta \overline{I} + \epsilon \beta \overline{C}) y_{1} \\ + (\beta y_{3} + \epsilon \beta y_{4}) y_{1}  -(\sigma+ \mu) y_{2} 
\end{split}
 \end{equation} 
 
  \begin{equation} \label{eq:48}
\frac{dy_{3}}{dt} = \sigma y_{2} - (\gamma+ \mu) y_{3} 
 \end{equation}
 
  \begin{equation} \label{eq:49}
\frac{dy_{4}}{dt} = \rho \gamma y_{3} - (\tau+ \mu) y_{4} 
 \end{equation}

 Let us define the Liapunov function as:
 
 \vspace{0.1 in}
 
 V : IR$^{4} \rightarrow IR^{+}$ by V(y) = $y_{1}^{2} $+ $y_{2}^{2} + y_{3}^{2} + y_{4}^{2}$ 
  \vspace{0.1 in}
 
Evaluating the condition on the parameters that make $\frac{dV(y)}{dt} \leq $ 0, $\forall y \in IR^{4}$ and if the directional derivative of
the solution y = y(t$_{0}, y_{0}$, t) of the IVP (Eq. \ref{eq:5} to \ref{eq:8}) along the closed curve V(y) is negative at any time t, then the endemic equilibrium point $\overline{x}$ of the original IVP is asymptotically stable.

  \begin{equation} \label{eq:50}
\frac{dV(y)}{dt} = \frac{\partial V(y)}{\partial y} . \frac{dy}{dt}
 \end{equation}  
 
  \begin{equation} \label{eq:51}
 \begin{split}
\frac{dV(y)}{dt} = 2 \big [ -(\beta y_{3} + \epsilon \beta y_{4})\overline{S} - (\beta y_{3} + \epsilon \beta y_{4}) y_{1} -\\ (\beta \overline{I}  + \epsilon \beta \overline{C} -(\omega+ \mu)) y_{1} - w(y_{2}+ y_{3} + y_{4})\big ] y_{1} \\
+ 2  \big [ (\beta y_{3} + \epsilon \beta y_{4})\overline{S} + (\beta \overline{I} + \epsilon \beta \overline{C}) y_{1} \\ + (\beta y_{3} + \epsilon \beta y_{4}) y_{1}  -(\sigma+ \mu) y_{2}  \big ]y_{2} \\
+ 2  \big [ \sigma y_{2} - (\gamma+ \mu) y_{3} \big ] y_{3} + 2  \big [ \rho \gamma y_{3} - (\tau+ \mu) y_{4} \big ]y_{4}
\end{split}
 \end{equation}

 \begin{equation} \label{eq:52}
 \begin{split}
\frac{dV(y)}{dt} = 2 \big [ -(\beta\overline{S} y_{1} y_{3} + \epsilon \beta \overline{S}y_{1} y_{4}) -(\beta y_{1}^{2} y_{3} + \epsilon \beta y_{1}^{2} y_{4})  \\ - (\beta \overline{I}  + \epsilon \beta \overline{C} -(\omega+ \mu)) y_{1}^{2}  -w(y_{1}y_{2}+ y_{1}y_{3} + y_{1}y_{4})\big ] \\ + 2 \big [ (\beta \overline{S}y_{2} y_{3} + \epsilon \overline{S} \beta y_{2}y_{4}) + (\beta \overline{I} + \epsilon \beta \overline{C}) y_{1}y_{2} + \\ (\beta  y_{3} + \epsilon \beta y_{4}) y_{1}y_{2} - (\sigma+ \mu) y_{2}^{2}  \big ] \\+ 2  \big [ \sigma y_{2} - (\gamma+ \mu) y_{3} \big ] y_{3} + 2  \big [ \rho \gamma y_{3} -(\tau+ \mu) y_{4}^{2} \big ]
\end{split}
 \end{equation}

  \begin{equation} \label{eq:53}
 \begin{split} 
  \frac{dV(y)}{dt}= 2 \big [ -\beta S y_{1} y_{3} -\epsilon \beta S y_{1} y_{4} - (\beta \overline{I}  + \epsilon \beta \overline{C}  -(\omega+ \mu)) y_{1}^{2} \\ -w(y_{1}y_{2}+ y_{1}y_{3} + y_{1}y_{4})\big ] + 2 \big [\beta S y_{2}y_{3} + \epsilon \beta S y_{2}y_{4} \\  + (\beta \overline{I} + \epsilon \beta \overline{C}) y_{1}y_{2} -(\sigma+ \mu) y_{2}^{2} \big ]  + 2  \big [ \sigma y_{3} y_{2} - (\gamma+ \mu) y_{3}^{2} \big ] \\ + 2  \big [ \rho \gamma y_{3} - (\tau+ \mu) y_{4}^{2} \big ]
  \end{split}
 \end{equation} 
 
 This above identity is including the variable S $\in \Omega$ and its limit value in the neighborhood of $\overline{x}$
is $\overline{S}$. Hence,

 \begin{equation} \label{eq:54}
 \begin{split} 
  \frac{dV(y)}{dt}= 2 \big [ -\beta \overline{S}y_{1} y_{3} -\epsilon \beta \overline{S} y_{1} y_{4} - (\beta \overline{I}  + \epsilon \beta \overline{C}  -(\omega+ \mu)) y_{1}^{2} \\ -w(y_{1}y_{2}+ y_{1}y_{3} + y_{1}y_{4})\big ] + 2 \big [\beta \overline{S}y_{2}y_{3} + \epsilon \beta \overline{S}y_{2}y_{4} \\  + (\beta \overline{I} + \epsilon \beta \overline{C}) y_{1}y_{2} -(\sigma+ \mu) y_{2}^{2} \big ]  + 2  \big [ \sigma y_{3} y_{2} - (\gamma+ \mu) y_{3}^{2} \big ] \\ + 2  \big [ \rho \gamma y_{3} - (\tau+ \mu) y_{4}^{2} \big ]
  \end{split}
 \end{equation} 

Using the property \textit{$\pm$ ab $\leq$  a$^{2}$ + b$^{2}$  , $\forall$a, b $\in$ IR} , yield the following inequality:

 \begin{equation} \label{eq:55}
 \begin{split} 
  \frac{dV(y)}{dt}  \leq  2 \big [ -\frac{1}{2}\beta \overline{S}(y_{1}^{2}+y_{3}^{2}) -\frac{1}{2} \epsilon \beta \overline{S}(y_{1}^{2}+y_{4}^{2}) \\ - (\beta \overline{I}  + \epsilon \beta \overline{C}  -(\omega+ \mu)) y_{1}^{2} -\frac{1}{2} w(3y_{1}^{2} +y_{2}^{2}+ y_{3}^{2} + y_{4}^{2})\big ] \\ + 2 \big [ \frac{1}{2}\beta \overline{S}(y_{1}^{2}+y_{3}^{2}) +\frac{1}{2} \epsilon \beta \overline{S}(y_{1}^{2}+y_{4}^{2}) +  \frac{1}{2}(\beta \overline{I}  + \epsilon \beta \overline{C}) (y_{1}^{2} + y_{2}^{2})\\   -(\sigma+ \mu) y_{2}^{2} \big ] \\ +  2 \big [ \frac{1}{2}  \sigma (y_{2}^{2}+ y_{3}^{2})  - (\gamma+ \mu) y_{3}^{2} \big ]  + 2  \big [\frac{1}{2} \rho \gamma (y_{3}^{2}+ y_{4}^{2}) - (\tau+ \mu) y_{4}^{2} \big ]
  \end{split}
 \end{equation}

 \begin{equation} \label{eq:56}
 \begin{split} 
  \frac{dV(y)}{dt}  \leq   -\big [ \beta \overline{S}(y_{1}^{2}+y_{3}^{2}) + \epsilon \beta \overline{S}(y_{1}^{2}+y_{4}^{2}) + \\ 2 (\beta \overline{I}  + \epsilon \beta \overline{C}  -(\omega+ \mu)) y_{1}^{2} + w(3y_{1}^{2} +y_{2}^{2}+ y_{3}^{2} + y_{4}^{2})\big ] \\ + \big [- \beta \overline{S}(y_{2}^{2}+ y_{3}^{2})-\epsilon \beta \overline{S}(y_{2}^{2}+ y_{4}^{2})  -( \beta \overline{I}+\epsilon \beta \overline{C})(y_{1}^{2}+ y_{2}^{2})\\ + 2 (\sigma+\mu) (y_{2}^{2}\big ] \\ + \big [- \sigma (y_{2}^{2}+ y_{3}^{2})+2 (\gamma+\mu) y_{3}^{2}\big ] \\ + \big [-\rho \gamma (y_{3}^{2}+ y_{4}^{2}) + 2 (\tau+ \mu)y_{4}^{2}\big ]
 \end{split}
 \end{equation}
The endemic equilibrium point $\overline{x}$ is asymptotically stable only if the RHS of the above inequality is positive definite. That is, every coefficient C$_{i}$ of the variable y$_{i}^{2}$, i = 1, 2, , 4 must be positive.

 \begin{equation} \label{eq:57}
 C_{1}=\beta \overline{S} + \epsilon \beta \overline{S} + 2 (\beta \overline{I}  + \epsilon \beta \overline{C}  -(\omega+ \mu)) + 3w-(\beta \overline{I}  + \epsilon \beta \overline{C}) \geq 0
 \end{equation}
 
 \begin{equation} \label{eq:58}
 C_{2}=w-\beta \overline{S}- \beta \overline{S}-(\beta \overline{I}  + \epsilon \beta \overline{C})+ 2(\sigma+ \mu) -\sigma \geq 0
 \end{equation}
 
  \begin{equation} \label{eq:59}
 C_{3}=\beta \overline{S}+ w - \beta \overline{S} -\sigma + 2(\gamma+ \mu) -\rho \gamma \geq 0
 \end{equation}

  \begin{equation} \label{eq:60}
 C_{4}=\epsilon \beta \overline{S} + w - \epsilon \beta \overline{S} -\rho \gamma + 2(\tau+ \mu) \geq 0
 \end{equation}
 
 It follows that the message propagation endemic equilibrium point is asymptotically stable if the following
conditions hold:

\begin{enumerate}
 \item  $ \beta (\overline{S}+ \overline{I})+ \epsilon \beta(\overline{S}+ 2 \overline{C})+ 2w-\mu \geq 0 $
  \item $ w- \beta (\overline{S}+ \overline{I})- \epsilon \beta(\overline{S}+ \overline{C})+ \sigma +2\mu \geq 0 $
 \item  $ w-\sigma+ 2(\gamma+\mu)-\rho \gamma \geq 0 $
 \item  $ w-\rho \gamma + 2(\tau+\mu)\geq 0 $
\end{enumerate}

To summarize, we have analyzed the stability of the message propagation equilibrium and endemic equilibrium point and proved that  message equilibrium and endemic equilibrium point  are asymptotically stable. A system may be stable or unstable (in our case) depending on the delay being finite or infinite. A system is asymptotically stable, if the  total delay in the system is less than infinity, unconditionally and does not depend on system parameters for the upper bound definition (marginal stability). Since we have shown that the system is asymptotically stable under conditions in message equilibrium and endemic equilibrium, it implies that the system will not experience infinite delay in processing and routing messages in DTN while endowed with a maximal coverage. This is a crucial theoretical guarantee that enables us to investigate the scenario from a practical point of view and ensures that the promising results in simulation and performance metric evaluation (as observed in section) are not by accident.

\begin{table}[ht]
\caption{Settings for numerical analysis}
\centering
\begin{tabular}{|l|c|c|}
\hline
\textbf{Parameter} &\textbf{ Value }  \\
 \hline
Transmission rate ($\beta$) &  1.0335e-005\\
 \hline
Transmission reducing factor ($\epsilon$) & 0.084 \\
\hline
 Probability of infected to become carrier ($\rho$) & 0.95 \\
 \hline
 Recovery rate  ($\gamma$) &  0.0714\\
\hline
Rate from susceptible to exposed class ($\sigma$) & 0.0714 \\
\hline
Rate at which immune decays ($\omega$) & 0.0588\\
 \hline
Rate from carrier to removal  ($\tau$) & 5.4795e-004
\\
\hline
Death rate ($\mu$) &  6.8493e-5 \\
\hline
Interval of integration of the model   & [0 - 2*365]\\
\hline
Initial susceptible value ($S_{0}$)  & 0.86\\
\hline
Initial carrier value  ($C_{0}$) & 0.03\\
\hline
Initial infected value ($I_{0}$)  & 0.02\\
\hline
Initial exposed value  ($E_{0}$) & 0.01\\
\hline
Initial recovered value  ($R_{0}$) & 0.08\\
\hline
\end{tabular}
\label{tab:settings_numerical_analysis}
\end{table}

\section{Numerical Analysis of CISER Model}

In this section, we will analyze the proposed CISER model numerically, with a set of input values for the parameters. Such numerical analysis will help finding  solutions for the differential equations used in CISER model and to match the solution with the simulation results. The values for the parameters are listed in table \ref{tab:settings_numerical_analysis}.

\begin{figure}[!t]
\centering
\includegraphics[width=3.5in]{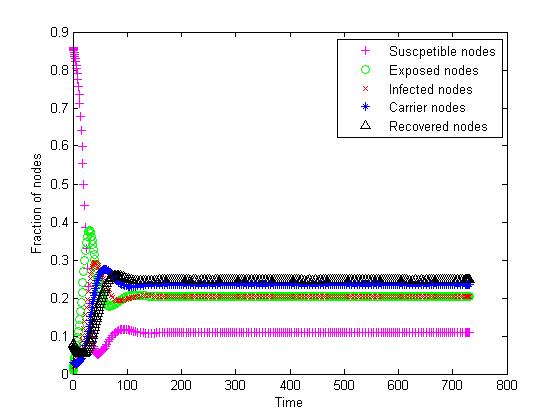}
% where an .eps filename suffix will be assumed under latex, 
% and a .pdf suffix will be assumed for pdflatex; or what has been declared
% via \DeclareGraphicsExtensions.
\caption{Data points of CISER model  }
\label{fig_sim_n}
\end{figure}

Applying the possible values of the parameters listed in table \ref{tab:settings_numerical_analysis}), in the equations \ref{eq:5}, \ref{eq:6}, \ref{eq:7}, \ref{eq:8} and \ref{eq:9},  we will get an array of values which are plotted in Fig. \ref{fig_sim_n}. Since our aim is to find a function \textit{f(x)} for representing fraction of population in each of the \textit{S}, \textit{E}, \textit{C}, \textit{I} and \textit{R} classes, from the set of data points generated, we have used cubic spline interpolation method.

Cubic spline interpolation divides the entire approximate interval to a set of sub-intervals and interpolate  using a different polynomial for each of them.  The results of cubic spline interpolation for  infected nodes is listed in  Fig. \ref{fig_sim_n2}. From the figure it can be observed that the interpolation perfectly matches  the set of input data points with value of error as zero, while approximating the accuracy. We  have also verified the accuracy of the fit  using \textit{Mean Squared Error} (MSE) method.

\begin{figure}[!t]
\centering
\includegraphics[width=3.5in]{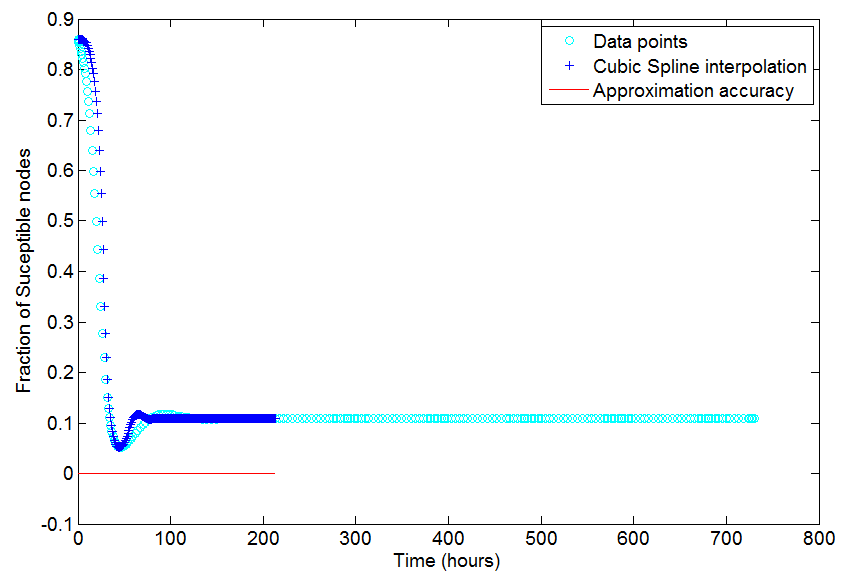}
% where an .eps filename suffix will be assumed under latex, 
% and a .pdf suffix will be assumed for pdflatex; or what has been declared
% via \DeclareGraphicsExtensions.
\caption{ Cubic spline interpolation of susceptible nodes }
\label{fig_sim_n1}
\end{figure}

%\begin{equation} \label{eq:121}
%MSE =\frac{1}{N} \sum_{1}^{N} (s_{i} -\hat{s_{i}})
% \end{equation}  

%Where \textit{N} is the size of the data array and $s_{i}$ is the original data points  %and  $\hat{s_{i}}$ is the prediction using interpolation.
 
\begin{figure}[!t]
\centering
\includegraphics[width=3.5in]{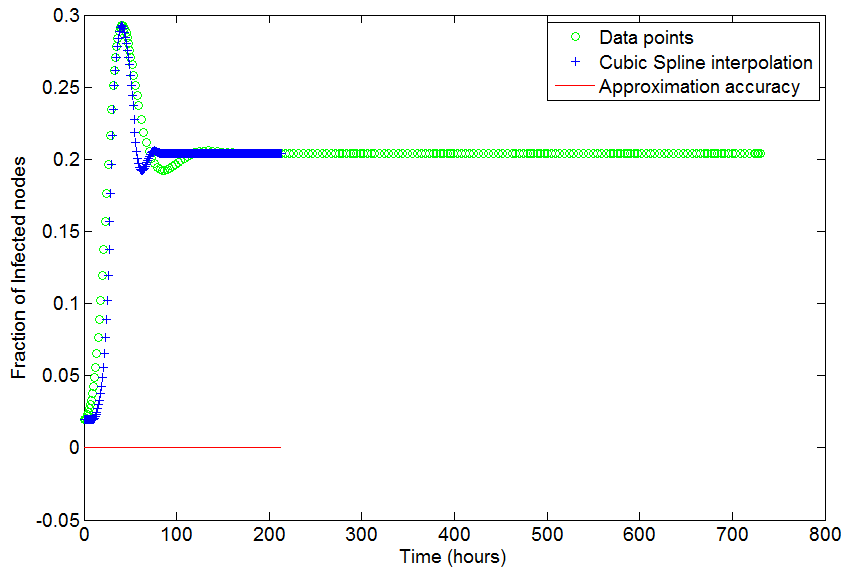}
% where an .eps filename suffix will be assumed under latex, 
% and a .pdf suffix will be assumed for pdflatex; or what has been declared
% via \DeclareGraphicsExtensions.
\caption{Cubic spline interpolation of infected nodes  }
\label{fig_sim_n2}
\end{figure}

The cubic spline function  \textit{C(x)} for the tabular  data, (x$_{1}$,y$_{1}$),  (x$_{2}$,y$_{2}$),... (x$_{n}$,y$_{n}$) is  represented  using the following equation.

 \begin{equation} \label{eq:1_22}
C(x) = p_{0} + p_{1}*x + p_{2}*x^{2} + p_{3}*x^{3}
 \end{equation}  

Where, $p_{0}, p_{1}, p_{2},$ and $p_{3}$ are constants. The cubic spline \textit{ C(x)}, is having following properties.

\begin{enumerate}

\item \textit{C(x)} is composed of cubic polynomial pieces C$_{k}$(x)
	
\textit{C(x) }= C$_{k}$(x), if \textit{x} $\in$ [x$_{k}$, x$_{k+1}$], \textit{k} = 1,2,..n-1.
	
\item C(x$_{k}$)= y$_{k}$, \textit{k} = 1,..n.	(interpolation)

\item C$_{k-1}$(x$_{k}$)= C$_{k}$(x$_{k}$), k = 2,..n-1.	

\item C$^{'}_{k-1}$(x$_{k}$)= C$^{'}_{k}(x_{k}$), k = 2,..n-1.

\item C$^{''}_{k-1}$(x$_{k}$)= C$^{''}_{k}(x_{k}$), k = 2,..n-1.

\end{enumerate}

With the data points we have generated, the cubic spline function \textit{C(t)}, is defined by the Eq. \ref{eq:1_22},  on the interval [0, 730], with not-a-knot end conditions. Using such equation, we found the solution of  differential equations for susceptible, exposed, infected, carrier and recovered fraction of nodes with values of constants obtained using cubic spline interpolation. As an example, for infected nodes,  the cubic spline interpolation in the interval  [728.34, 726.697] with coefficients, p$_{3}$=0, p$_{2}$=0, p$_{1}$=0.0052 and p$_{0}$=0.108, gives, 

 \begin{equation} \label{eq:1_27}
\textit{C(t)} = 0.1087 + 0.0052*t
 \end{equation}

The error of any interpolation is the maximum difference between original function and approximated function. The error bound for not-a-knot cubic spline interpolation, \textit{C(t)} over the interval [t$_{0}$, t$_{N}$], can be expressed using following equation.

 \begin{equation} \label{eq:1_28}
|| f(t) -C(t)|| \leq  h^{4} \frac{5}{384}  || f^{4}(t)||_{t_{0},t_{N}}
 \end{equation}  
 
Where the width \textit{h} =max$_{i} |x_{i}-x{i-1}|$. So the error associated with not-a-knot cubic spline interpolation is O(h$^{4}$). For instance, for infected nodes, C(t) in the cubic spline interpolation in the interval  [0.0461, 0.0921], with coefficients a$_{3}$=0.4891, a$_{2}$=-0.09, a$_{1}$=0.003 and a$_{0}$=0.02, gives 
 \begin{equation} \label{eq:1_29}
C(t) = 0.02+ 0.0031*(t) - 0.09*(t)^{2} + 0.4891*(t)^{3} 
 \end{equation}
We obtain the value of error as 1.7258e-007 by applying  Eq. \ref{eq:1_28}. This is reasonably accurate by any standards and justifies the use of splines as approximation tool.

\section{ SIMULATION OF CISER MODEL }

In this section, we will discuss the performance evaluation of the  proposed CISER model  in  DTN environment and compare the results with SIR model using ONE Simulator \cite{keranen2009one}, which is a well-known DTN simulator.

\subsection{Simulation setup} \label{setup}

We have used \textit{Random Way Point} (RWP) mobility model as part of the simulation using synthetic mobility traces. We have considered a network of 160 nodes  moving randomly in a closed region of 8 $\times$ 8 Km, for 12 hours. Each node is having 2 MB buffer space to store a message of size 500 KB to 1 MB, which is having a Time To Live (TTL) value of 300 minutes. The transmission speed of the nodes is 250 kbps with a node speed of 4-10 km/hr and a transmission range of 100 meters. Since our CISER model is developed for analyzing message propagation in Epidemic DTN, we have used Epidemic routing as the basic routing protocol. We assume that, at time $t=0$, a single source node (\textit{S}) exists in the network having a message to be delivered to a single destination (\textit{D}).

We have implemented the proposed CISER model on  Epidemic routing   and compared the routing performance with SIR model of Epidemic routing. For convenience, we label the protocols as Epidemic-CISER and Epidemic-SIR. The values of equations obtained using numerical analysis is represented by the legend Epidemic-CISER-Approximation. The simulation settings are shown in Table \ref{tab:settings1}. 

\begin{table}[ht]
\caption{Simulation settings}
\centering
\begin{tabular}{|l|c|c|}
\hline
\textbf{Parameter} &\textbf{ Value }  \\
\hline
 Simulation time & 43200s=12h\\
 \hline
 Number of nodes & 160 \\
\hline
 Transmission Speed of nodes & 250 Kbps \\
 \hline
 Time To Live (TTL)  & 300 minutes \\
 \hline
Message Creation Interval & 25-30 seconds\\
\hline
Message Size & 500-1024 KB \\
\hline
 Wait Time & 10-30 seconds\\
 \hline
Device buffer& 2 MB\\
\hline
Initial energy & 4800 mAh \\
\hline
Scan energy & 1 mAh \\
\hline
Receive energy & 4 mAh \\
\hline
Transmit energy & 4 mAh \\
\hline
\end{tabular}
\label{tab:settings1}
\end{table}

\subsection{Adaption of parameters in DTN environment}

In order to perform the simulation in a DTN environment, we need to adapt the parameters used in CISER model to DTN characteristics and functionality. Since we have used SIR epidemic model for comparison with CISER model, we will explain how we set the parameters for both SIR model and CISER model below.

\subsubsection{SIR model}

Revisiting the equations used for  SIR model in Epidemic DTN, 

\begin{equation} \label{eq:55}
\frac{dS}{dt} =-\beta I 
 \end{equation} 

\begin{equation} \label{eq:56}
\frac{dI}{dt} =-\beta I - \gamma I
 \end{equation} 

\begin{equation} \label{eq:57}
\frac{dR}{dt} =\gamma I
 \end{equation} 

The parameter $\beta$ represents the contact (infection) rate among DTN nodes and $\gamma$ represents the recovery rate  (rate of depletion), which is the rate at which either the messages are delivered or discarded (due to expiry of TTL). Referring to \cite{groenevelt2005message}, the contact rate ($\beta$) among nodes in an Epidemic DTN environment can be calculated as follows

\begin{equation} \label{eq:34}
 \beta \approx \frac{2*\omega*r*E[V]}{A^{2}}
 \end{equation}

The constant $\omega$ is specific to the RWP mobility model and having a value 1.3683. The area of closed region is represented by \textit{A} and \textit{r} represents the transmission range of the nodes in the network. The term \textit{E[V]} represents the  average relative speed between two nodes in the network.  Applying the values  for \textit{A}, $\omega$, \textit{r} and \textit{E[V]} in equation \ref{eq:34}, from our simulation settings (discussed above), the value of $\beta$ is approximately 1.03335 $\times 10^{-5}$. Regarding recovery rate ($\gamma$), for very large TTL, it is very rare to discard a message before delivery, so a node is said to be recovered, if it  delivers the message directly to the destination, which yields $\beta$ =  $\gamma$.

\subsubsection{CISER model}

In the case of CISER model, the contact rate, $\beta$ is calculated using the same equation \ref{eq:34}. Since a susceptible node is infected when it is in the communication range of an infected node,  $\lambda$ is the same as the contact rate $\beta$. The same scenario applies for both $\sigma$ and $\gamma$. We  assume the existence of noise ($\epsilon$) in the neighbor discovery, which will reduce the contact rate to a fraction $\epsilon\beta$ and  is set to a value 0.084.

We have set the initial energy of the node as 4800 mAh During message transfer, nodes depletes 1 mAh for neighbor discovery and 4 mAh for both message transmission and reception. If the remaining energy level of a node   falls below a value of 5 mAh, we assume such node as a dead node. Since the initial energy level of a node is quite high, death of nodes is very rare to happen and we assume $\tau$ as 5.4795$\times 10^{-4}$.

We assume that an infected node is being a carrier with a probability $\rho$ as 0.95. The reason for such an assumption is that due to flooding-based data forwarding approach used in Epidemic routing, a node with a message copy (infected)  is having a higher chance to forwards the message further in the network (thereby becoming a carrier). Also, a node is said to be recovered, if it either  delivers the message to the destination or discarded the message due to expiry of TTL. Since we assume a large TTL value for the message, the only way for recovery when the node is in contact with a destination node, which means $\omega$ = $\beta$.

\subsection{Performance metrics}

We use \textit{delivery ratio}, \textit{overhead ratio} and \textit{delivery delay} as the metrics for evaluation of our proposed  CISER model in epidemic DTN environment. These metrics are defined as follows.
 
$\bullet$	\textbf{Delivery Ratio (DR)}: It is the ratio of total number of messages delivered  to the total number of messages generated at the nodes for routing.

$\bullet$	\textbf{Overhead Ratio (OR):} It is  the ratio of difference between the total number of relayed messages ($N_{r}$) and the total numbers of delivered messages ($N_{d}$) to the total number of delivered messages. It is defined as

\begin{equation}\label{eq:p2}
OR = \frac{(\Sigma N_{r} - \Sigma N_{d})}{\Sigma N_{d}}
\end{equation}

$\bullet$	\textbf{Delivery Delay (DD):} It is the average time taken for a message to reach the destination.

\subsection{Results using synthetic traces}

 \begin{figure}[!t]
\centering
\includegraphics[width=3.0in]{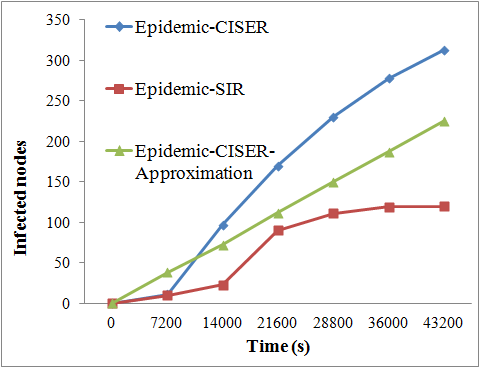}
% where an .eps filename suffix will be assumed under latex, 
% and a .pdf suffix will be assumed for pdflatex; or what has been declared
% via \DeclareGraphicsExtensions.
\caption{Propagation of infection with time }
\label{fig_2}
\end{figure}

In this section, we will discuss the results of the simulation CISER model using synthetic traces.

%\begin{figure}[!t]
%\centering
%\includegraphics[width=2.5in]{Results_Available_new}
% where an .eps filename suffix will be assumed under latex, 
% and a .pdf suffix will be assumed for pdflatex; or what has been declared
% via \DeclareGraphicsExtensions.
%\caption{Number of Susceptible nodes for different simulation runs varying total number of nodes in the network }
%\label{fig_1}
%\end{figure}

We have analyzed the total number of infected nodes in both SIR and CISER model for different simulation runs varying simulation time from 0 to  43200 seconds (12 hours) for a total number of 480 nodes in the network.  We have also calculated the approximate number of infected nodes using the equation \ref{eq:1_27} obtained from the result of cubic spline interpolation and is represented by the legend "Epidemic-CISER-Approximation".

The figure \ref{fig_2} shows that for smaller duration (upto 7200 s), the total number of infected nodes are the same for both SIR and CISER models. However, as simulation progresses, the number of infected nodes are decreasing for SIR model because of the large overhead (See Fig. \ref{fig_sim5}) associated with message forwarding.

\begin{figure}[!t]
\centering
\includegraphics[width=3.0in]{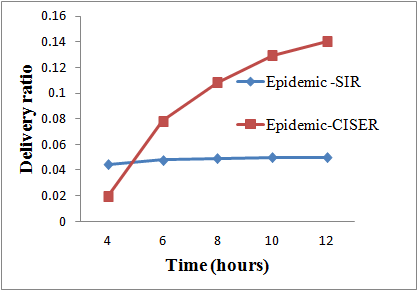}
% where an .eps filename suffix will be assumed under latex, 
% and a .pdf suffix will be assumed for pdflatex; or what has been declared
% via \DeclareGraphicsExtensions.
\caption{Comparison of delivery ratio }
\label{fig_sim4}
\end{figure}

\begin{figure}[!t]
\centering
\includegraphics[width=3.0in]{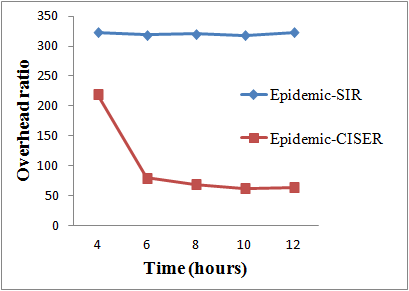}
% where an .eps filename suffix will be assumed under latex, 
% and a .pdf suffix will be assumed for pdflatex; or what has been declared
% via \DeclareGraphicsExtensions.
\caption{Comparison of overhead ratio }
\label{fig_sim5}
\end{figure}

\begin{figure}[!t]
\centering
\includegraphics[width=3.0in]{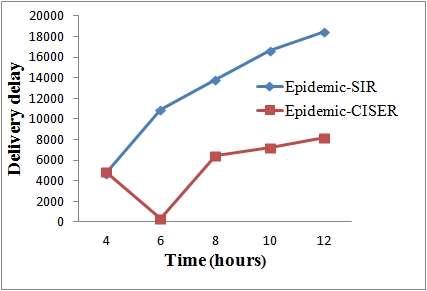}
% where an .eps filename suffix will be assumed under latex, 
% and a .pdf suffix will be assumed for pdflatex; or what has been declared
% via \DeclareGraphicsExtensions.
\caption{Comparison of delivery delay}
\label{fig_sim6}
\end{figure}

For analyzing the DTN routing performance metrics such as delivery ratio, delivery delay and overhead ratio, we have varied the simulation time from four hours to twelve hours. Fig. \ref{fig_sim4} depicts the comparison of delivery ratio of both SIR and CISER models. From the figure, it can be observed that although, the SIR model achieves a higher delivery ratio (for smaller simulation duration), as simulation time increases CISER model achieves higher delivery ratio compared to SIR model. This is because, in case of CISER model, not all infected nodes are propagating the message, instead those having enough resources are acting as carriers, thereby  improving the overall delivery ratio.

Similarly, comparing the overhead ratio as in Fig. \ref{fig_sim5}, CISER model achieves a significant reduction (about 80 $\%$) in overhead ratio compared to SIR model. This is because of large number of message forwarding involved in SIR model, as a source/relay node floods the message to all its neighbor nodes, which results in resource overhead, as the nodes having limited storage space and energy. However,  with CISER model, resource-constrained nodes are not involved in message forwarding, which will reduce the overall overhead.

Regarding the delivery delay (Fig.\ref{fig_sim6}), both SIR model and CISER models achieves lowest delivery delay initially, but as simulation progresses, CISER model achieves almost 35 $\%$ reduction in in delivery delay compared to SIR model. As  number of infected (carrier nodes) increases, messages are having high chance of delivering to the destination.

\subsection{Results using real-world traces }

We have used real-world traces, such as Infocom \cite{chaintreau2007impact} and MIT Reality \cite{eagle2006reality}  for analyzing the performance of the CISER model in real time.

\begin{figure}[!t]
\centering
\includegraphics[width=3.0in]{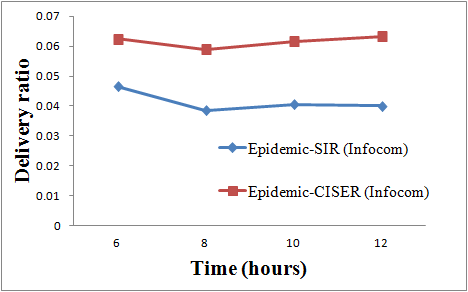}
% where an .eps filename suffix will be assumed under latex, 
% and a .pdf suffix will be assumed for pdflatex; or what has been declared
% via \DeclareGraphicsExtensions.
\caption{Delivery ratio comparison using Infocom trace}
\label{fig_sim8}
\end{figure}

\begin{figure}[!t]
\centering
\includegraphics[width=3.0in]{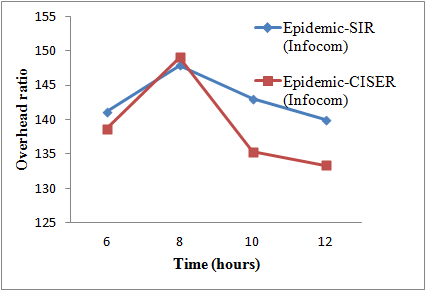}
% where an .eps filename suffix will be assumed under latex, 
% and a .pdf suffix will be assumed for pdflatex; or what has been declared
% via \DeclareGraphicsExtensions.
\caption{Overhead ratio comparison using Infocom trace}
\label{fig_sim9}
\end{figure}

\begin{figure}[!t]
\centering
\includegraphics[width=3.0in]{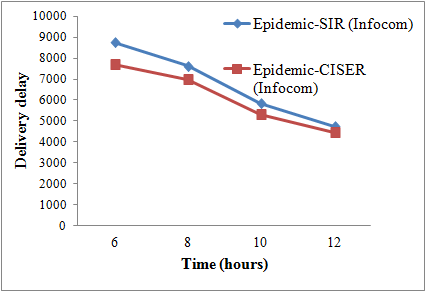}
% where an .eps filename suffix will be assumed under latex, 
% and a .pdf suffix will be assumed for pdflatex; or what has been declared
% via \DeclareGraphicsExtensions.
\caption{Delivery delay comparison using Infocom trace}
\label{fig_sim10}
\end{figure}

The Infocom trace consists of  Bluetooth sightings of 41 participants of the INFOCOM 2005 conference carrying iMotes for four days (March 7-10, 2005), in Grand Hyatt Miami. The iMotes periodically scan the neighbourhood at the interval of every two minutes. The comparison of delivery ratio, overhead ratio and delivery delay of SIR and CISER models of Epidemic routing is listed in Fig. \ref{fig_sim8},  Fig. \ref{fig_sim9} and Fig. \ref{fig_sim10} respectively. Compared to synthetic data sets, the delivery ratio is much lower using real-world traces. This is because of the inherent limitations of real-world traces, such as low population analyzed and low sensing interval. However, CISER model achieves almost 49 $\%$ gain in  delivery ratio compared to SIR model. Comparing the overhead ratio, overhead ratio of CISER model is slightly less compared to SIR model. This is because the number of messages generated is much less  compared to the synthetic model, so there is not much variations in number of message forwarding in both the models. Regarding delivery delay, CISER model achieves almost 10 $\%$ reduction compared to SIR model.

\begin{figure}[!t]
\centering
\includegraphics[width=3.0in]{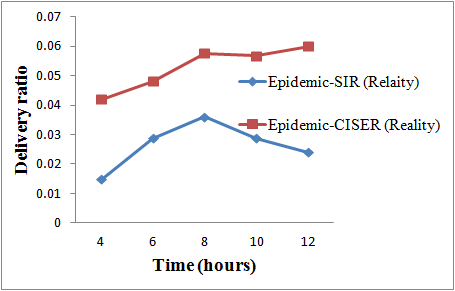}
% where an .eps filename suffix will be assumed under latex, 
% and a .pdf suffix will be assumed for pdflatex; or what has been declared
% via \DeclareGraphicsExtensions.
\caption{Delivery ratio comparison using Reality trace }
\label{fig_sim11}
\end{figure}

\begin{figure}[!t]
\centering
\includegraphics[width=3.0in]{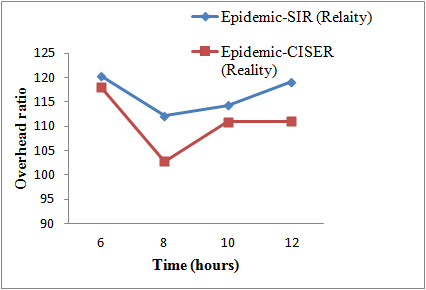}
% where an .eps filename suffix will be assumed under latex, 
% and a .pdf suffix will be assumed for pdflatex; or what has been declared
% via \DeclareGraphicsExtensions.
\caption{Overhead ratio comparison using Reality trace }
\label{fig_sim12}
\end{figure}

\begin{figure}[!t]
\centering
\includegraphics[width=3.0in]{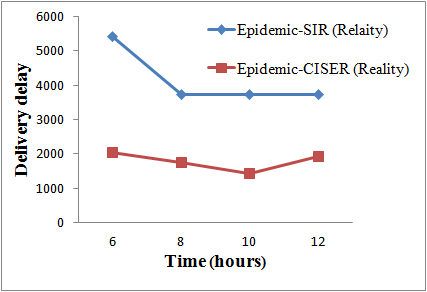}
% where an .eps filename suffix will be assumed under latex, 
% and a .pdf suffix will be assumed for pdflatex; or what has been declared
% via \DeclareGraphicsExtensions.
\caption{Delivery delay comparison using Reality trace }
\label{fig_sim13}
\end{figure}

MIT Reality contains human mobility traces collected during an experiment at MIT campus with 100 students and staffs for the academic year 2004-05. Each user is equipped with Nokia 6600 mobile phone for a period of 9 months. The generated data consist of about 450,000 hours of information, which represents user's location, communication and usage behavior. The comparison of delivery ratio for SIR and CISER  using Reality trace is plotted in Fig. \ref{fig_sim11}, Fig. \ref{fig_sim12} and Fig. \ref{fig_sim13} respectively. Similar to the results using Infocom trace, delivery ratio of  CISER model is higher than SIR model and the overhead ratio is slightly lower than SIR model. However, there is a significant reduction (about 56 $\%$) in delivery delay for CISER model, compared to the results using Infocom trace.

%\begin{figure}[!t]
%\centering
%\includegraphics[width=3.0in]{Results_SIR_CISER_CAMBRIDGE_DR}
% where an .eps filename suffix will be assumed under latex, 
% and a .pdf suffix will be assumed for pdflatex; or what has been declared
% via \DeclareGraphicsExtensions.
%\caption{Delivery ratio comparison using Cambridge data set}
%\label{fig_sim9}
%\end{figure}

In summary, we have conducted simulations for validating the proposed CISER model using both synthetic and real-world traces in DTN environment and the results highlight that the CISER model achieves better routing performance compared to the basic SIR model.

\section{Conclusion}

Due to the time varying network topology and resource constraints of nodes in DTN, providing mathematical modeling for message propagation is a challenging task. Some of the researchers attempted to model the epidemic message propagation in DTN using the SIR model developed to study infectious  disease propagation pattern in human population. In this paper, we  extended the SIR model and proposed a novel CISER model for epidemic message propagation in DTN, with additional states to represent resource-constrained behavior of DTN nodes. Our contributions are summarized as follows. 

 We proposed CISER, a novel and efficient mathematical model for  representing the epidemic message forwarding characteristics in DTN based on  ameobiasis disease propagation patterns.  We have performed a qualitative study of the CISER model to identify the determinant parameter for the spread of the messages throughout the network. We have also analyzed the endemic equilibrium of epidemic message propagation and identified scenarios that may take place when message propagation dynamics is at its endemic steady state. We have performed a stability analysis of epidemic message propagation and proved that the message propagation is asymptotically stable.

 We have performed a numerical analysis of the proposed CISER model by varying the parameters used in the model to study the dynamics of epidemic message propagation. Finally, we have conducted extensive simulations with synthetic and real-world traces to evaluate the routing performance of the CISER model in a DTN environment. The results show that the CISER model achieves better routing performance compared to the basic SIR model in terms of higher delivery ratio, lower delivery delay and lower overhead ratio.


  \section*{Acknowledgments}
%\else
This work is partially supported by the Alexander von Humboldt Foundation through the post-doctoral research fellowship of one of the authors.
  %\section*{Acknowledgment}
%\fi

%\bibliographystyle{IEEEtran}
\bibliographystyle{IEEEtran}
\bibliography{IEEEabrv,impact_of_selfishness_epidemic}

\begin{IEEEbiography}[{\includegraphics[width=1in,height=1.15in,clip,keepaspectratio]{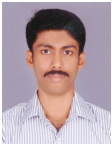}}]{Sobin C C}
Sobin CC received the BTech degree in Information Technology from the College of Engineering, Thalassery, Kerala (affiliated to Cochin University, Kerala), in 2004, and the M.Tech degree in Computer Science and Engineering from Indian Institute of Technology(IIT), Madras in 2010. He is currently doing his PhD in Computer Science from Indian Institute of Technology(IIT), Roorkee. He worked as an Assistant Professor in MES Engineering College, Kerala, before joining for PhD in IIT Roorkee. His research interests include routing in Delay Tolerant Networks, mathematical modeling, Internet-of Things. He is a student member of the IEEE.
\end{IEEEbiography}

%\vspace{-10 mm}
\begin{IEEEbiography}[{\includegraphics[width=1in,height=1.15in,clip,keepaspectratio]{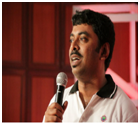}}]{Snehanshu Saha}
Snehanshu Saha holds Masters Degree in Mathematical Sciences at Clemson University and Ph.D. from the Department of Mathematics at the University of Texas at Arlington in 2008. He is a Professor of Computer Science and Engineering at PESIT South since 2011 and heads the Center for Applied Mathematical Modeling and Simulation. He has published 40 peer-reviewed articles in International journals and conferences and been IEEE Senior member and ACM professional member since 2012. Snehanshu’s current and future research interests lie in Data Science, Machine Learning and applied computational modeling. 
\end{IEEEbiography}
%\vspace{-10 mm}

\begin{IEEEbiography}[{\includegraphics[width=1in,height=1.15in,clip,keepaspectratio]{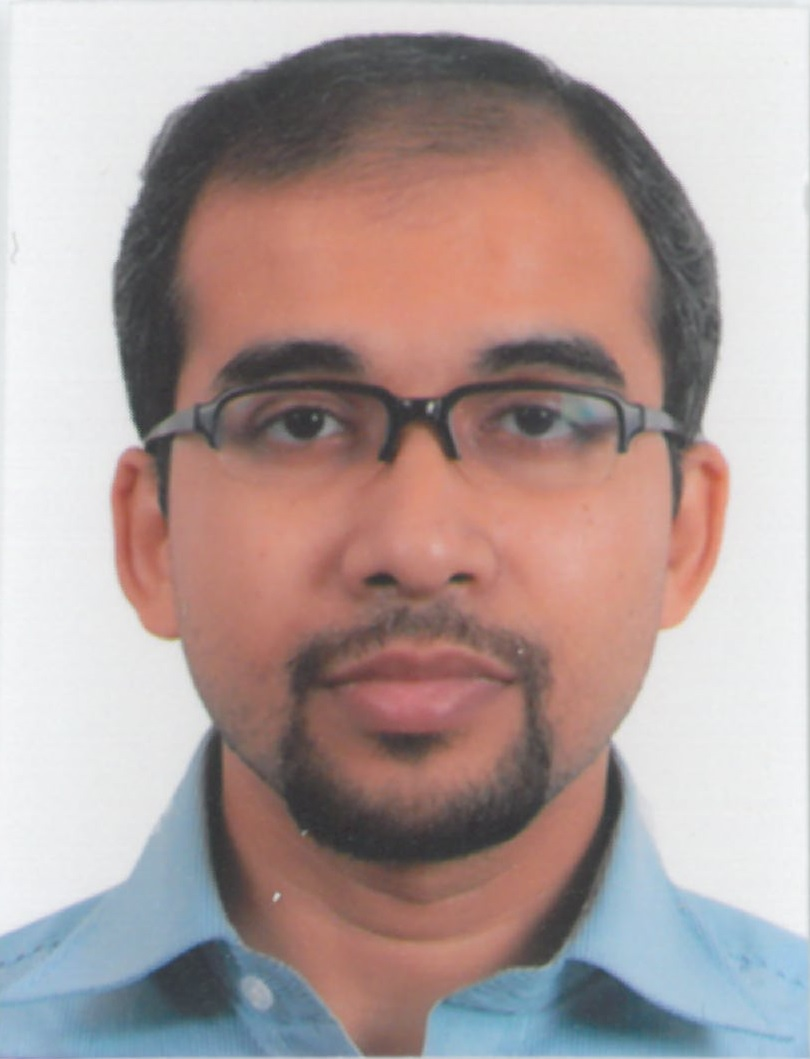}}]{Vaskar Raychoudhury}

Vaskar Raychoudhury is currently working as an Alexander von Humboldt Post-doctoral Research Fellow jointly with the Universität Mannheim and Technische Universität Darmstadt, Germany. He received his PhD in Computing from The Hong Kong Polytechnic University in 2010 and went to join Institut Telecom SudParis, in France to work as a post-doctoral research fellow. In 2011 he joined Department of Computer Science and Engineering, Indian Institute of Technology (IIT) Roorkee as an Assistant Professor. His research interests include mobile and pervasive computing and networking, Internet-of-Things, Wireless Sensor Networks and Big Data management. He keeps publishing high-quality journals and conferences in these areas. He has served as program committee member in Globecom, ICDCN, and reviewers of top IEEE transactions and Elsevier journals. He is a member of ACM, and a senior member of IEEE.
\end{IEEEbiography}

\begin{IEEEbiography}[{\includegraphics[width=1in,height=1.15in,clip,keepaspectratio]{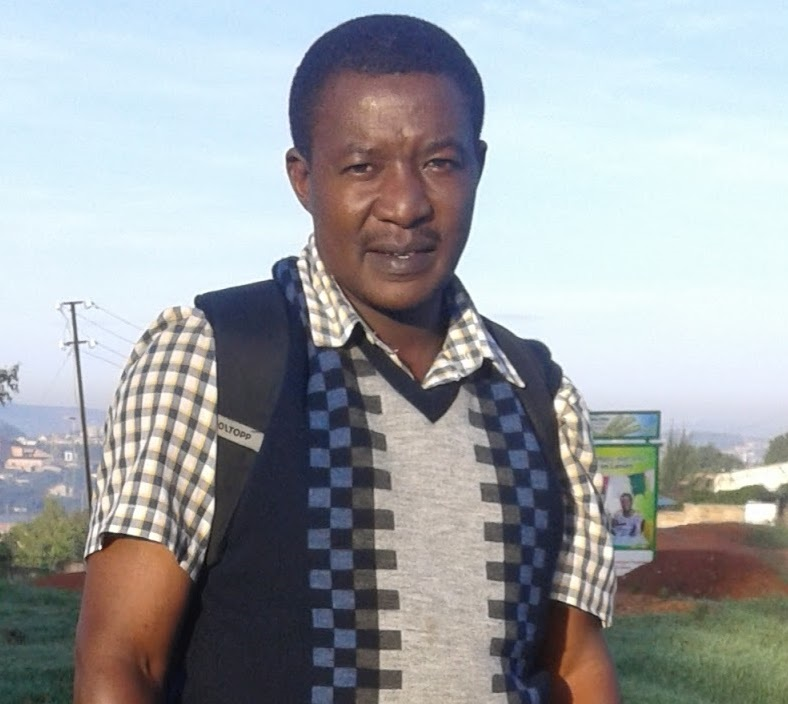}}]{Hategekimana   Fidele }

Hategekimana Fidele holds a Masters degree in Mathematics from Bangalore University since 2009 and he is a PhD scholar since September 2013 at Jain University. He has been a lecturer of Applied mathematics at Adventist University of Central Africa and a visiting lecturer of the same course in different private universities  and institutions of higher learning in Rwanda. He is interested  in Mathematical Biology  and Mathematical modelling of infectious diseases. He is the author of 3 papers in peer review journals and  has held four local conferences.
\end{IEEEbiography}
\end{document}